\documentclass{aa}  

\usepackage{keyval}
\usepackage{graphicx}
\usepackage{natbib}
\usepackage{xcolor}
\usepackage[normalem]{ulem}
\bibpunct{(}{)}{;}{a}{}{,}

\usepackage{txfonts}

\usepackage[colorlinks=false]{hyperref}

\begin{document}

   \title{Towards an Observational Detection of Halo Spin Bias \\using Spin-Orbit Coherence}

   \author{Yigon Kim
          \inst{1}
          \and
          Antonio D. Montero-Dorta
          \inst{2}
          \and
          Rory Smith
          \inst{3, 4}
          \and
          Jong-Ho Shinn
          \inst{5}
          }

   \institute{Astronomy program, Department of Physics and Astronomy, Seoul National University, 
   1 Gwanak-ro, Gwanak-gu, Seoul, Republic of Korea, 08826.\\
              \email{yigon28@snu.ac.kr}
         \and
             Departamento de F\'isica, Universidad T\'ecnica Federico Santa Mar\'ia, 
Avenida Vicu\~na Mackenna 3939, San Joaqu\'in, Santiago, Chile\\
             \email{amonterodorta@gmail.com}
        \and
             Departamento de F\'isica, Universidad T\'ecnica Federico Santa Mar\'ia, 
Avenida Espa\~na 1680, Valpara\'iso, Chile\\
             \email{rorysmith274@gmail.com}
        \and
             Millenium Nucleus for Galaxies (MINGAL)
        \and
        Korea Astronomy and Space Science Institute (KASI), 776 Daedeokdae-ro, Yuseong-gu, Daejeon, Republic of Korea, 34055.
             }

   \date{Received Febuary 10, 2025; accepted }
 
  \abstract
  % context heading
   {
   The clustering of dark-matter halos depends primarily on halo mass. However, at fixed halo mass, numerical simulations have revealed multiple secondary dependencies. This so-called secondary halo bias has important implications for our understanding of structure formation and observational cosmology. Despite its significance, the effect has not yet been measured observationally with statistical confidence.
   }
  % aims heading
   {
    We aim to develop the first observational method to probe halo spin bias: the secondary dependence of halo clustering on halo spin at fixed halo mass.
   }
  % methods heading
   {
    We use a proxy for halo spin based on the coherent motion of galaxies within and around a halo. This technique is tested using the IllustrisTNG hydrodynamical simulation and subsequently applied to a group catalog from the Sloan Digital Sky Survey (SDSS). By splitting the SDSS groups according to this spin proxy and measuring the two-point correlation function of the resulting samples, the existence of halo spin bias is investigated.
   }
  % results heading
   {
   We find consistent indications that, at fixed mass, groups with higher values of the spin proxy exhibit higher bias than those with lower spin proxy values, on scales of 5–15 $h^{-1}$$\mathrm{Mpc}$. The highest significance is seen for groups with halo masses $M_{\rm h} \gtrsim 10^{13.2}$ $h^{-1}\rm M_\odot$, for which 85$\%$ of the sampled measurements display the expected trend. As we continue to improve the method, our results could open new avenues for studying the connection between halo spin and the large-scale structure with upcoming spectroscopic surveys.
   }
   {}


   \maketitle

\section{Introduction}

In the so-called non-linear regime of structure formation, overdensities of dark matter (DM) can decouple from the expansion of the Universe and continue collapsing to form DM halos. Within these collapsed objects, which constitute the building blocks of the large-scale structure of the Universe (LSS), galaxies form by cooling and condensation of gas inside the halos' potential wells (\citealt{White1978, White1991}). In this context, linear halo bias describes the relation between the density contrast of DM halos and that of the underlying matter density field on large scales, i.e, $b_{\mathrm h} = \delta_{\mathrm h}/\delta_{\mathrm m}$ (e.g., \citealt{Mo1996}). It can be easily shown that this parameter also indicates the relative clustering strength of both components, as $b_{\mathrm h}^2 = \xi_{\mathrm h}/\xi_{\mathrm m}$, where $\xi_{\mathrm h}$ and $\xi_{\mathrm m}$ are the auto-correlation functions of halos and matter. 

It is well known that halo bias depends strongly on the internal properties of halos. Halo mass is responsible for the primary dependence, as a direct manifestation of the more fundamental dependence of halo bias on the peak height of density fluctuations, $\nu$. More massive halos are more tightly clustered than less massive halos, as described by the $\Lambda$-cold dark matter ($\Lambda$-CDM) structure formation formalism (e.g., \citealt{Press1974,ShethTormen1999,Sheth2001,ShethTormen2002}). At fixed halo mass, a number of additional {\it{secondary dependencies}} have been unveiled using mostly cosmological simulations (see, e.g., \citealt{Sheth2004,gao2005,Wechsler2006,Gao2007,Angulo2008,2008Li,faltenbacher2010, Lazeyras2017,2018Salcedo,han2018,Mao2018, SatoPolito2019, Johnson2019, Ramakrishnan2019,MonteroDorta2020B,Tucci2021,MonteroDorta2021_mah,MonteroRodriguez2024, MonteroDorta2025}). As an example, lower mass halos that assemble a significant portion of their mass early on are more tightly clustered than those that form at later times, an effect referred to as {\it{halo assembly bias}}.

Another secondary halo bias effect that has been discussed in the literature is the so-called {\it{spin bias}}, the dependence on halo spin, $\lambda$ (e.g., \citealt{Gao2007,SatoPolito2019,Johnson2019,Tucci2021, MonteroDorta2021_SZ, MonteroDorta2021_hsb}).
Halo spin is commonly defined in simulations as a dimensionless parameter proportional to the total angular momentum of the DM particles \citep{Peebles1969,Bullock2001}, namely: 
\begin{equation}
    \lambda = \frac{|J|}{\sqrt{2} \, M_{\text{vir}} V_{\text{vir}} R_{\text{vir}}},
\end{equation}
for the Bullock et al. version, 
where $J$ is the halo angular momentum inside a sphere of radius $R_{\text{vir}}$ and mass $M_{\text{vir}}$, and $V_{\text{vir}}$ is its circular velocity at virial radius $R_{\text{vir}}$. The effect of spin bias on the clustering of halos can be divided into two regimes. At the low-mass end, lower-spin halos are more tightly clustered than higher-spin halos of the same mass, due to the effect of splashback halos \citep{Tucci2021}. At the high-mass end, the opposite trend is observed. It appears that this {\it{intrinsic}} dependence could be related to fundamental theories that link the angular momentum of halos to the tidal field (e.g. such as the Tidal Torque Theory \citealt{BarnesEfstathiou1987}). The physical origins of spin bias, and secondary bias in general, are, however, yet to be established (see, e.g., \citealt{Dalal2008, Paranjape2018, MonteroDorta2025}).

Although secondary halo bias has been extensively characterized from cosmological simulations, its existence is yet to be established observationally. The attempts to probe the effect can be divided in two groups. Some works focused on demonstrating the halo bias effect directly from observations, using proxies for halo properties (e.g.,\citealt{Miyatake2016,Lin2016,MonteroDorta2017B,Niemiec2018,Sunayama2022}). Other works addressed the manifestation of secondary halo bias on the galaxy population, an effect that is generally called galaxy assembly bias (see, e.g., \citealt{Obuljen2020,Salcedo2022,Wang2022} -- note that different definitions of this effect are assumed). Despite some claims, it is fair to say that a broadly accepted proof for the existence of secondary bias has not been presented, due to a variety of issues including selection and contamination effects, the small amplitude of the signal for assembly bias in some ranges, or the inherent challenge of measuring halo masses, to name but a few.

In this work, we investigate, for the first time, halo spin bias using observational data. At face value, spin bias provides a major advantage as compared to halo assembly bias, in terms of a potential detection of secondary bias. Unlike assembly bias, the spin-bias signal increases in amplitude towards the high-mass end (e.g., \citealt{Tucci2021}), which allows the use of clusters to probe the effect. Conversely, however, measuring the spin of halos has been historically very challenging \citep{Mroczkowski2019}. \cite{MonteroDorta2021_SZ} presented a proof of concept analysis on the feasibility of the rotational kinetic Sunyaev Zel'dovich (rkSZ) effect as the basis of a potential observational probe for halo spin bias. Although the correlation between the rkSZ signal and halo spin in hydrodynamical simulations is indeed sufficient to produce secondary bias, the observational aspects of the rkSZ measurement represent still a major obstacle \citep{Mroczkowski2019}.

Our observational probe for spin bias is based on the coherent orbital motion of neighboring galaxies around a target halo. This proxy has been already measured observationally in \citet{JLee2019a, JLee2019b}. The spin direction of the target galaxy is found in these works to correlate with the averaged motion of galaxies around the target galaxy. Using simulations, \citet{YKim2022b} evaluated the factors that determine the strength of this coherence signal, finding that it increases with the mass of the target halo and its spin, $\lambda$. Thus, by measuring the motions of neighboring galaxies about a target halo, an estimate for the magnitude and direction of its spin vector can be obtained. We refer to this coherence as the `Spin-Orbit coherence', or SOC, herein. We also refer to the magnitude of the coherence motion as the `coherence signal' for simplicity. The observational probe presented here opens up the doorway for an observational analysis of halo spin bias, which could provide an important validation of the $\Lambda$CDM cosmological model.

The paper is structured as follows. In Sect. \ref{sec:data}, we introduce the data employed in this work, including both the observational and simulation data. Sect. \ref{sec:method} describes the methodology that we follow, with especial emphasis on our spin proxy based on Spin-Orbit coherence. In Sect. \ref{sec:result}, we present the main result of the paper, our measurement of spin bias. Sect. \ref{sec:dcuss} is devoted to discussing the robustness and significance of our measurement. Finally, in Sect. \ref{sec:concl}, we summarize our main conclusions and discuss the implications of our results. Throughout this work, we adopt the standard $\Lambda$CDM cosmology \citep{planck2016}, with parameters $\Omega_{\mathrm m} = 0.3089$, $\Omega_{\mathrm b} = 0.0486$, $\Omega_\Lambda = 0.6911$, $H_0 = 100\,h\, {\mathrm km\, s^{-1}Mpc^{-1}}$ with $h=0.6774$, $\sigma_8 = 0.8159$, and $n_s = 0.9667$.

\section{Data} \label{sec:data}

Our analysis is based on publicly available data from both a hydrodynamical simulation, which we use to test the feasibility of our spin proxy, and a spectroscopic survey, where the spin-bias probe is applied. In this section, we describe the data and the selection criteria for both datasets.

\subsection{Illustris TNG300} \label{sec:tng300}

Part of our analysis is based on data from the Illustris-TNG magneto-hydrodynamical cosmological simulation (hereafter TNG for simplicity, \citealt{Pillepich2018b,Pillepich2018,Nelson2018_ColorBim,Nelson2019,Marinacci2018,Naiman2018,Springel2018}). The TNG simulation suite was produced using the {\sc arepo} moving-mesh code \citep{Springel2010} and is considered an improved version of the previous Illustris simulation \citep{Vogelsberger2014a, Vogelsberger2014b, Genel2014}. The updated TNG sub-grid models account for star formation, radiative metal cooling, chemical enrichment from SNII, SNIa, and AGB stars, and stellar and super-massive black hole feedback. These models were calibrated to successfully reproduce a set of observational constraints that include the observed $z=0$ galaxy stellar mass function, the cosmic SFR density, the halo gas fraction, the galaxy stellar size distributions, and the black hole -- galaxy mass relation (we refer the reader to the aforementioned papers for more information).

We analyze the largest box available in the database, TNG300-1 (hereafter TNG300\footnote{\url{https://www.tng-project.org/data/docs/specifications/}}), which provides an obvious advantage when it comes to measuring large-scale halo/galaxy clustering. TNG300 spans a side length of $205\,\,h^{-1}$Mpc and includes periodic boundary conditions. The TNG300 run followed the dynamical evolution of 2500$^3$ DM particles of mass $4.0 \times 10^7$ $h^{-1} {\mathrm M_{\odot}}$ and (initially) 2500$^3$ gas cells of mass $7.6 \times 10^6$ $h^{-1} {\mathrm M_{\odot}}$. 

In order to ensure mass resolution, we have imposed mass cuts of $M_{\mathrm h} > 10^{11}\, \, h^{-1}\mathrm{M}_{\odot}$ and  $M_{\rm sh} > 10^{9} \, \, h^{-1}\mathrm{M}_{\odot}$, for the total mass of bound particles in halos and subhalos, respectively\footnote{For reference, these cuts correspond to $\sim$2500 and 25 DM particles, respectively.}. These selection criteria reduce the sample to $254,382$ halos and more than $10^7$ subhalos. 

Since we aim to compare simulation and observational data directly, we have also created a mock catalog from TNG300. This mock catalog was constructed by converting the $z=0$ TNG300 snapshot into a 2D projected catalog, placing the observer at the center of the box. It contains positions, masses, velocities and the $R_{200}$ radii for halos, along with positions, masses and velocities for subhalos (galaxies). These properties are required to measure the average motion of galaxies around each halo. This mock catalog is used as a reference to evaluate the impact of projection effects on the measurement of spin bias and to aid in the optimization of certain parameters of our spin proxy. However, we note that it is not fully comparable to the SDSS in terms of neighbor mass distributions, owing to intrinsic differences between the TNG simulation and our SDSS selection criteria. The SDSS selection will be discussed in the following section.

\subsection{The SDSS LowZ group catalog} \label{sec:sdss}

\begin{figure}
    \centering
    \includegraphics[width=\columnwidth]{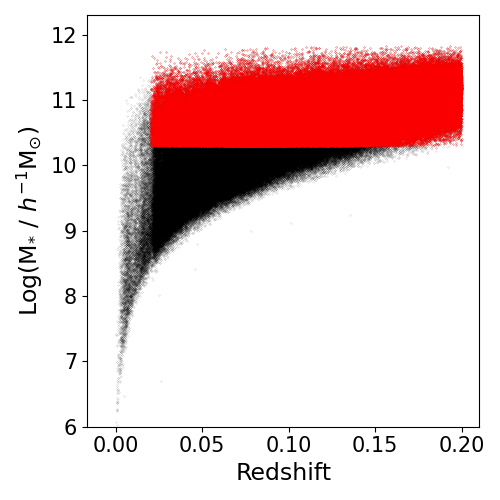}
    \caption{Stellar mass ($M_{*}$) versus redshift for galaxies in the SDSS LowZ catalog. Black markers show the entire sample while the red markers show our volume-limited sample.}
    \label{fig:sdssgal}
\end{figure}

The observational data employed in this work are part of the Sloan Digital Sky Server Data Release 13 (SDSS DR13, \citealt{SDSSLowZ}), which provides spectroscopy for galaxies throughout a large cosmological volume. These two aspects are crucial for our analysis -- the coherence measurement requires the precision of spectroscopic redshifts, while the clustering determination needs large cosmological volumes. The SDSS groups are extracted from the group catalog of \citet{Lim2017}, hereafter referred to as the ``SDSS LowZ group catalog", which adopts a redshift limit of $z < 0.2$ and the {\it{r}}-band magnitude limit of the legacy survey ($r < 17.77$). Briefly, the group-finding algorithm proceeds in an iterative way, first estimating the mass of the halo according to the stellar mass of its central galaxy, following an abundance-matching approach. Then, the group's size and velocity dispersion are evaluated in order to select the members. For a complete description of the method, we refer the reader to \citet{Yang2005, Lim2017}. 

Importantly, we use the halo mass-size relation from \citet{Pasquali2019} to determine the virial radius($R_{200}$) of each group, i.e.,

\begin{equation}
    R_{200}[h^{-1} kpc] = \frac{258.1 * M_{\mathrm{h,*}}^{1/3} * (\Omega_{m}/0.25)^{1/3}}{1+z}, 
\end{equation}

\noindent
where $M_{\mathrm{\rm h,*}}$ is the halo mass in units of $10^{12}$ $h^{-1}\rm{M_\odot}$ and $\Omega_{m}$ is the total matter density in the Universe.

The secondary bias dependencies, including spin bias, are known to display significant redshift evolution, due to the intrinsic dependence on peak height (e.g., \citealt{Gao2007, Tucci2021}). In order to minimize this effect and to avoid issues related to incompleteness, we employ volume-limited samples, as shown in Fig. \ref{fig:sdssgal}). In particular, we impose a stellar mass cut of $M_{*} > 10^{10.3} \ h^{-1} \mathrm{M}_{\odot}$, which ensures that the sample is complete in the redshift range $0.02 < z < 0.2$ - this is the sample that we use throughout this work (red dots in Fig. \ref{fig:sdssgal}). The stellar mass is estimated from the {\it{r-}}band luminosity and the {\it{g - r}} color following the prescription of \cite{Bell2003}. This selection produces a sample of a total of $410,042$ galaxies. 

Finally, it is important to impose a minimum number of group members that guarantees robustness in our SOC (Spin-Orbit coherence) measurements, while ensuring high-number statistics for our group sample. This is achieved by considering only groups with at least 3 members. We have checked that increasing this threshold to, e.g., 5 members, would results in significantly noisier results, as the number of groups decreases considerably. The total number of groups in our catalog after imposing this cut is $22,456$.

\section{Method} \label{sec:method}

\subsection{Spin-Orbit coherence} \label{sec:halocoh}

\begin{figure*}
    \centering
    \includegraphics[width=170mm]{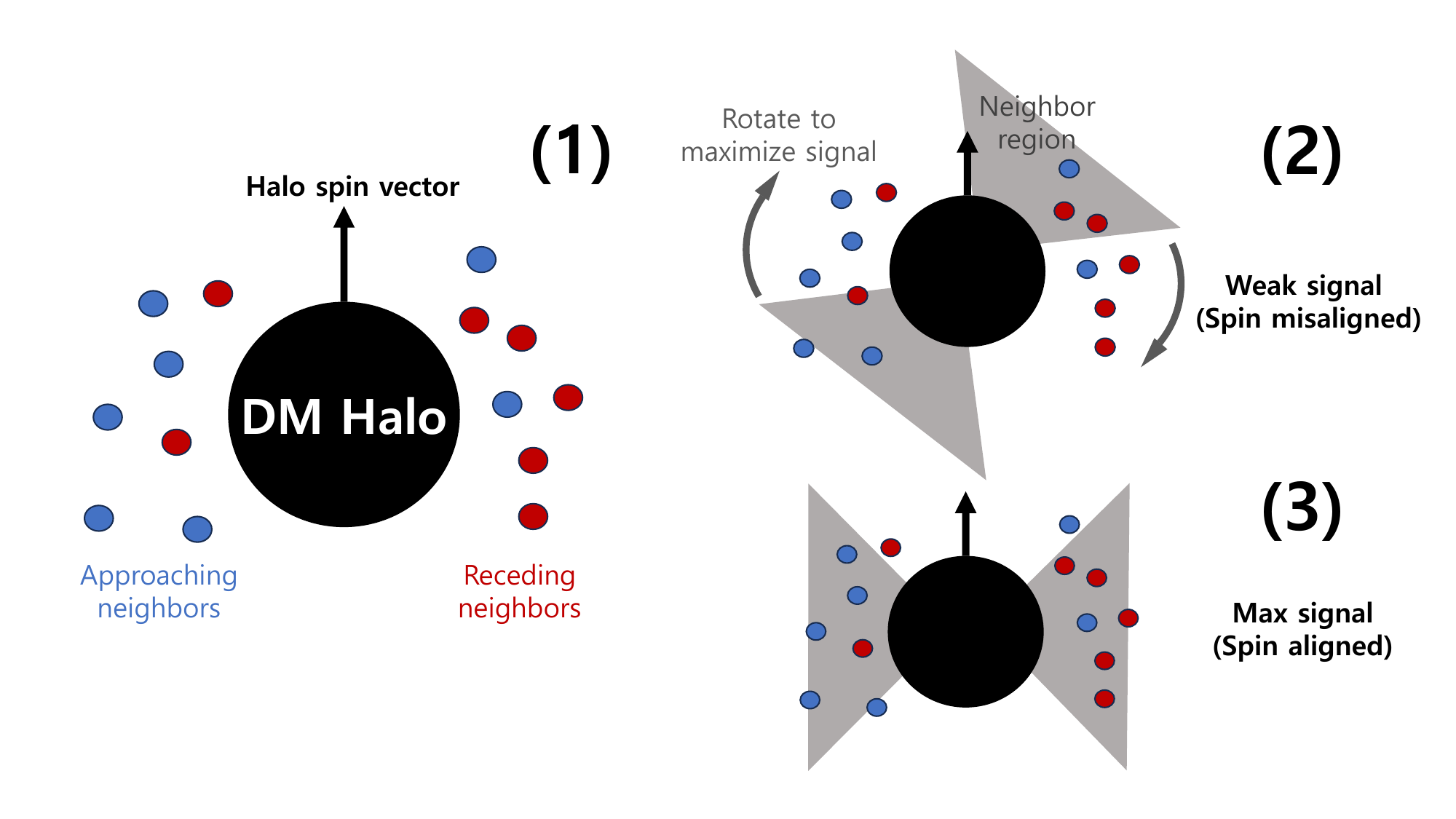}
    \caption{A simplified illustration of the concept behind spin proxy based on Spin-Orbit Coherence and its measurement. Cartoon 1 depicts the Spin-Orbit Coherence itself, while Cartoons 2 and 3 illustrate how the spin direction is determined by identifying the direction in which the coherence signal is strongest.}
    \label{fig:cartoon}
\end{figure*}

Our spin proxy based on the SOC is described in this section. The first evidence of this type of coherence was reported by \citet{JLee2019a}, using the Calar Alto Legacy Integral Field spectroscopy Area (CALIFA) survey \citep{califa1, califa2, califa3}. In that work, the spin of individual galaxies was measured using Integrated Field Unit (IFU) data. Further work, presented in \citet{JLee2019b}, showed the existence of a clear correlation between the spin direction of a central galaxy and the peculiar motion of the surrounding neighbor galaxies. The motion of the neighbor galaxies tends to align with the spin vector of their central galaxy. The parameter dependencies of this correlation in terms of the DM component were measured using cosmological simulations in \citet{YKim2022b}. In particular, \citet{YKim2022b} revealed a strong relation between the halo spin parameter, $\lambda$, and the strength of the coherence of neighbor motions. 

Here, we measure the coherence signal by taking a mass weighted mean of the tangential velocities of neighboring galaxies, determined with respect to the central (host) halo\footnote{Throughout this work, the central halo refers to the host halo, while neighboring galaxies can be either satellites of that host or other galaxies in its vicinity.}. These velocities are subsequently normalized by the virial velocity of the central halo, since naturally more massive halos tend to have higher neighbor velocities. By doing so, we are able to mitigate the intrinsic dependence of the coherence signal on the host halo mass. This is desirable because the halo spin parameter is by construction not strongly halo-mass dependent \citep{Peebles1969,Bullock2001}.

An illustration of the procedure to measure the spin proxy is shown in Fig. \ref{fig:cartoon}. We only use galaxies that fall within the shaded x-shaped region to calculate the signal. The opening angle of the x-shaped region was set to 45 degrees, following previous works \citep{JLee2019a, JLee2019b}. The aim is to exclude galaxies which would not be expected to provide much signal, as they are close to the spin vector describing the rotation of the central halo. After the projected X-cut is made about the halo, the mass weighted average of the line-of-sight velocity on either side of it is subsequently calculated. Galaxies on one side of the X-region will tend to move towards us (negative velocities) while on the other side they will move away from us (positive velocities), and this contributes to the strength of the spin proxy. 

Importantly, the region for identifying neighbor galaxies is set to be 1 $R_{200}$ on the sky, and the velocity equivalent of 1 $R_{200}$ along the line of sight assuming purely Hubble flow. This is a relatively narrow velocity cut that naturally excludes some cluster members. Our analysis using the TNG mock reveals that a narrow velocity cut produces a stronger signal, perhaps because we only consider the most relaxed members of the cluster in the calculation of the coherence signal. In addition, a narrow velocity cut helps mitigate the impact of external environmental effects by accidental inclusion of neighboring galaxies at larger distances. This potential contamination could artificially impact the signal. We will come back to this point in Sect. \ref{sec:statresult}, where we quantify the effect of varying the parameters of the fiducial model on the spin bias measurement.

Observationally, we do not know the spin direction of the galaxy groups/clusters. So, to try to determine its direction, we first set an arbitrary direction for the halo spin vector and measure the coherence signal. Then, we smoothly turn it around in steps of $10$ degrees through $360$ degrees, and calculate the coherence signal at each step. Our best choice for the direction of the spin vector is the step with the greatest value of coherence signal. The method to find the spin vector is shown in diagrams 2 and 3 of Fig. \ref{fig:cartoon}. We tested several choices of angle step size from 10 to 30 degrees, and we found that this choice does not significantly affect our results.

Using TNG300, we tested several conditions to maximize the correlation signal between $\lambda$ and the SOC. We found that this correlation is most sensitive to the host halo's mass, becoming significant for the range $M_{\mathrm h} > 10^{12} \, \, h^{-1}\mathrm{M}_{\odot}$. On the other hand, we checked that the mass of the neighbors does not significantly affect the signal (see also \citealt{YKim2022b}). 

\begin{figure}
    \centering
    \includegraphics[width=\columnwidth]
    {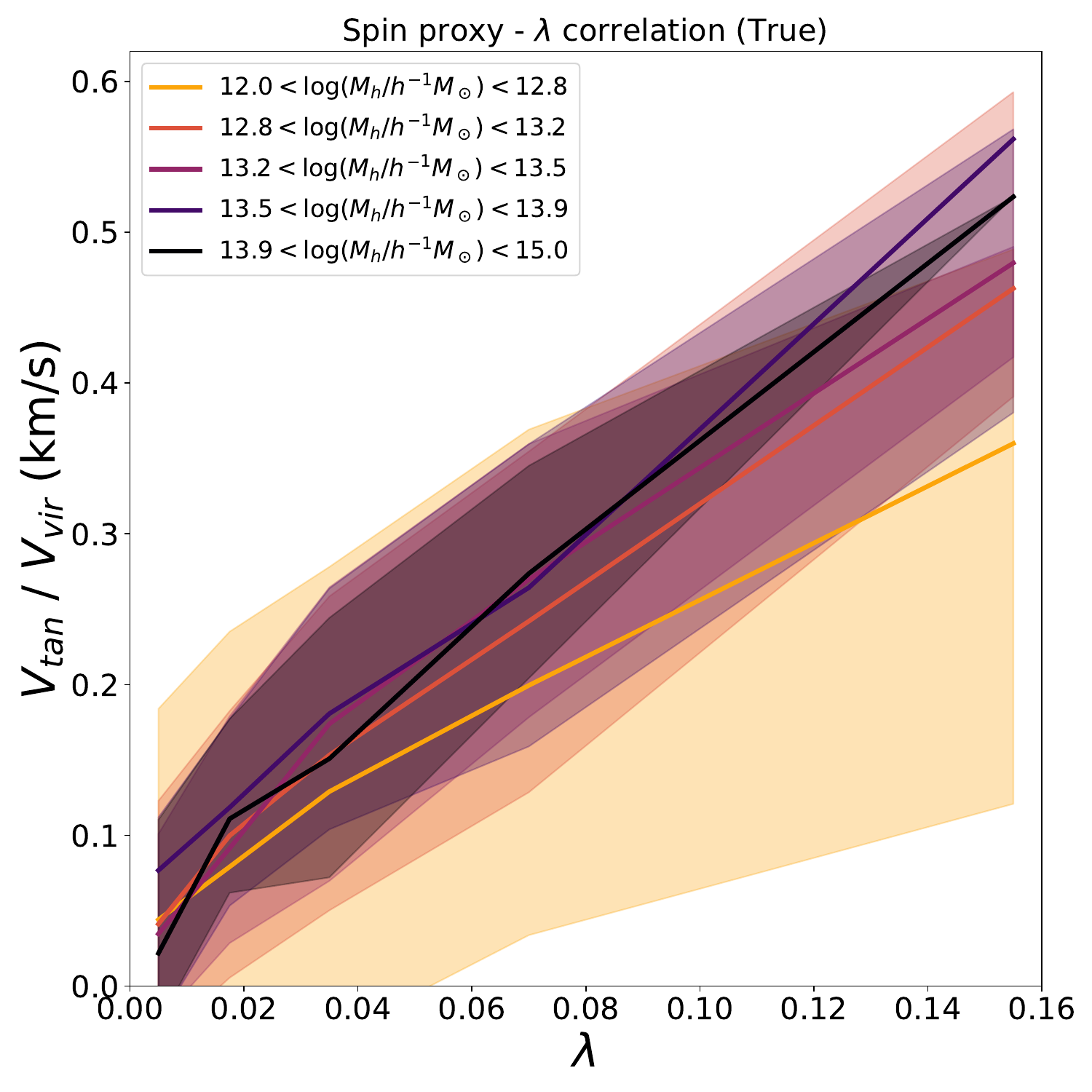}
    \caption{The correlation between the spin proxy and the true halo spin ($\lambda$) for several host halo mass bins in TNG300. The solid lines show the mean of the distribution, while the shaded regions indicate the interquartile range (25th to 75th percentiles). These measurements have been performed in 3D.}
    \label{fig:spincorr}
\end{figure}

Finally, the correlation between the true spin and the spin proxy that motivates this analysis is shown in Fig. \ref{fig:spincorr}, using 3D measurements from TNG300. Note that there is a clear correlation above $M_{\mathrm{h}} = 10^{12}$ $h^{-1}\mathrm{M}_{\odot}$, which becomes stronger towards the high-mass end.

\subsection{The spin bias measurement} \label{sec:halobi}

\begin{figure}
    \centering
    \includegraphics[width=\columnwidth]{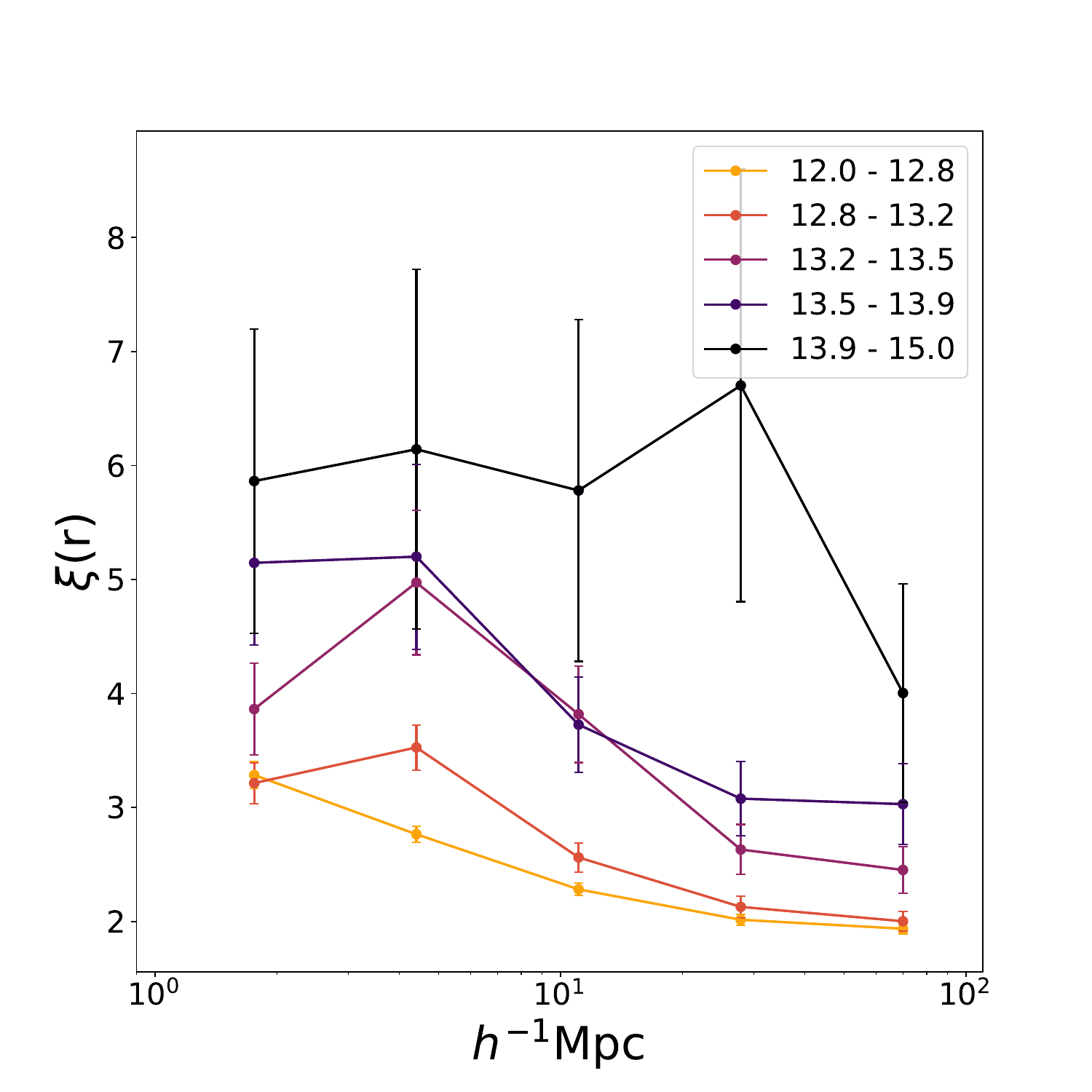}
    \caption{The cross correlation function of the SDSS LowZ group catalog as a function of group mass. The group mass range ($\log(M_{\mathrm{h}} / h^{-1}\mathrm{M}_{\odot})$) of each measurement is indicated in the legend. As predicted from theory, more massive groups are more strongly clustered, which is the primary dependency of halo bias.}
    \label{fig:firstbias}
\end{figure}

The measurement of halo spin bias, i.e., the secondary dependence of halo bias on spin at fixed halo mass, is performed, both in simulations and observations, following the standard procedure for secondary bias based on the ratio of correlation functions (see, e.g., \citealt{2018Salcedo, SatoPolito2019, MonteroDorta2020B}). In essence, we compute the correlation function for percentile subsets of halos based on their $\lambda$/spin proxy, in bins of halo mass. As mentioned before, the main dependence of halo bias is known to be virial mass, as a direct manifestation of the more fundamental dependence on the peak height of density fluctuations (e.g., \citealt{Sheth2001}). Fig. \ref{fig:firstbias} displays the auto-correlation function for groups as a function of mass in the SDSS, revealing this well-known relation: more massive groups have progressively higher clustering amplitude. This is an important sanity check for our analysis, which gives us confidence that the secondary dependencies at fixed halo mass can be measured observationally. Note that the mass binning adopted in Fig. \ref{fig:firstbias} will be also employed throughout in our spin-bias SDSS measurement (Sect. \ref{sec:sdssresult}).     

The relative bias for a subset of halos based on $\lambda$ (or, equally, the spin proxy), with respect to the entire halo population within a certain mass bin, i.e., $b_{\lambda}(r, M_{\text{h}})$, can be computed as follows:

\begin{equation}
    b_{\lambda}(r, M_{\text{h}}) = \frac{\xi_{\lambda}(r, M_{\text{h}})}{\xi_{tot}(r, M_{\text{h}})},
\end{equation}

\noindent where $\xi_{\lambda}(r, M_{\text{h}})$ is the cross-correlation between the corresponding ($\lambda$/spin proxy) percentile subset in that bin and the entire halo sample, and $\xi_{tot}(r, M_{\text{h}})$ is the cross-correlation between the whole mass bin and the entire halo sample (see \citealt{MonteroDorta2020B} for more details). A difference between the relative bias of higher- and lower-spin (or spin proxy) subsets at fixed halo mass would give rise to the secondary bias effect. Here, we use percentile subsets encompassing the upper and lower fractions of the mass bin (50-50$\%$ subsets). The cross-correlation functions are estimated applying the traditional \citet{1993ApJ...412...64L} method. The calculation is done using the CORRFUNC code \citep{corrfunc} and the relative bias is averaged on scales $5-15 h^{-1}$Mpc \citep{MonteroDorta2020B}. For the randoms, we use 10 times more objects than the data, also mimicking the geometry/volume of the data set (in the case of the SDSS, this includes mimicking the SDSS redshift distribution for the volume-limited sample). 

\begin{table}
\centering
\begin{tabular}{c||c}
\hline
Host halo (group) mass & $M_{\rm h} > 10^{12} \, \, h^{-1}\mathrm{M}_{\odot}$ \\
Neighbor stellar mass & $M_{\rm *} > 10^{10.3} \, \, h^{-1}\mathrm{M}_{\odot}$ \\
Neighbor range & $0.1-1R_{200,\rm halo}$ \\
Correlation function range & $5-15 h^{-1}$Mpc \\
Group member & $\ge 3$ \\
\hline
\end{tabular}
\caption{Fiducial configuration for the spin bias measurement in the SDSS.} \label{tab:config}
\end{table}

An important aspect in the determination of secondary bias is the error computation, both in the SDSS and in TNG300. Two different ways of estimating errors are employed in this work, depending on the secondary parameter used, for the sake of consistency. In TNG300, when $\lambda$, the real spin value, is used to split the halo population at fixed halo mass, errors are estimated using a traditional jackknife technique, since statistics allow it. In this case, the TNG300 simulation box is divided into $8$ equal-volume sub-boxes, and $8$ different spin-bias measurement are obtained by excluding only one sub-box at a time. The errors correspond to the standard deviation of this set of measurements. In the SDSS LowZ group catalog, employing a jackknife error estimation is problematic due to low number statistics.  For this reason, errors on the spin proxy measurements are determined using a bootstrap technique for each mass bin. A total of 100 bootstrap resampling subsets are generated, from which the mean and standard deviation of the resulting correlation functions are obtained. For consistency, this methodology is replicated in TNG300 whenever the spin proxy is employed, which helps us compared results between the observational data and the simulation.

In order to ensure the robustness of our measurement, two additional checks were performed in the SDSS group catalog. First, we verified that the redshift distributions of the entire group sample, as well as the higher- and lower-spin subsets, are similar in each mass bin. Second, since the distribution of halo masses is not uniform across the mass range of interest, we set the binning so that each bin contains an equal number of halos. This scheme is adopted for all measurements.

Finally, the set of parameters used for probing spin bias in the SDSS is summarized in Table \ref{tab:config}. 

\section{Results} \label{sec:result}

In this section, we use our spin proxy to probe for spin bias observationally. We first test our method on the TNG300 mock in order to evaluate the feasibility of the measurement. Subsequently, the method is applied to the SDSS LowZ group catalog and the robustness and significance of the resulting measurement are evaluated.

\subsection{Testing the spin proxy in the simulation} \label{sec:tngresult}

\begin{figure*}
    \centering
    \includegraphics[width=170mm]{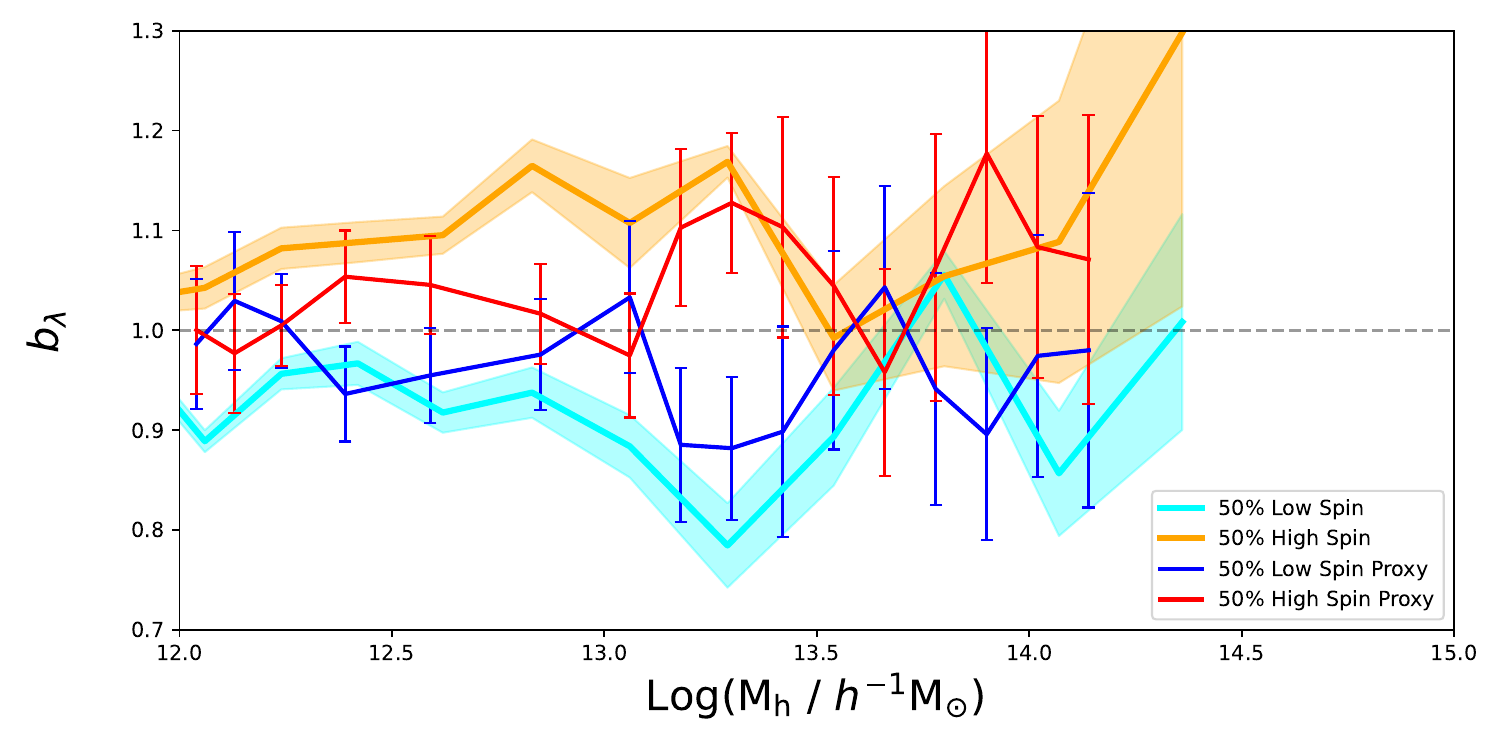}
    \caption{Halo spin bias measured in projection from the TNG300 mock. Higher spin and higher spin proxy measurements are shown in orange and red solid lines/symbols, while the lower spin and lower spin proxy measurements are represented by cyan and blue solid lines, respectively. See text for information regarding the computation of errors.}
    \label{fig:tng300bias}
\end{figure*}

To demonstrate that the spin proxy based on Spin-Orbit Coherence (SOC) is a reliable tracer of halo spin, we first tested it using simulations. As shown in \citet{MonteroDorta2020B}, the TNG300 simulated halo population displays a significant halo spin bias, among other secondary dependencies of halo clustering. To evaluate the feasibility of our spin proxy, we first performed the halo spin bias measurement in the TNG300 mock catalog. These results are shown shown in Fig. \ref{fig:tng300bias}. Each subset of halos corresponds to the 50$\%$ upper (orange) and lower (cyan) half of the $\lambda$ distribution in each mass bin. The 1-$\sigma$ jackknife error of the relative bias is plotted as shaded regions, estimated using a set of sub-volumes, as described in Sect. \ref{sec:halobi}. Despite the fact that we are using projected quantities in order to match the observations, the spin bias signal still remains, i.e., for all mass bins, the orange lines are above the cyan lines. It is noteworthy that when measured from a sufficiently large simulation box, the signal for $\lambda$ is known to continue increasing with halo mass (see, e.g., \citealt{SatoPolito2019}), which is not clearly seen here at the highest masses due to low-number statistics. 

After we verified that the spin bias signal remains when the measurement is performed in projection, we repeated the analysis in the mock using the spin proxy instead of the $\lambda$ parameter. These results are represented by the red and blue points in Fig. \ref{fig:tng300bias}. Again, we use the 50$\%$ highest (red) and lowest (blue) spin-proxy values. We recall also that, when the spin proxy is used to estimate secondary bias, errors are calculated by bootstrapping. Fig. \ref{fig:tng300bias} demonstrates that the spin proxy shows similar qualitative behavior in terms of spin bias as the $\lambda$-based measurement. This gives us confidence that the proxy can be successfully applied to observations, as we discuss in the following section.

\subsection{Measuring spin bias in observations} \label{sec:sdssresult}

\begin{figure*}
    \centering
    \includegraphics[width=170mm]{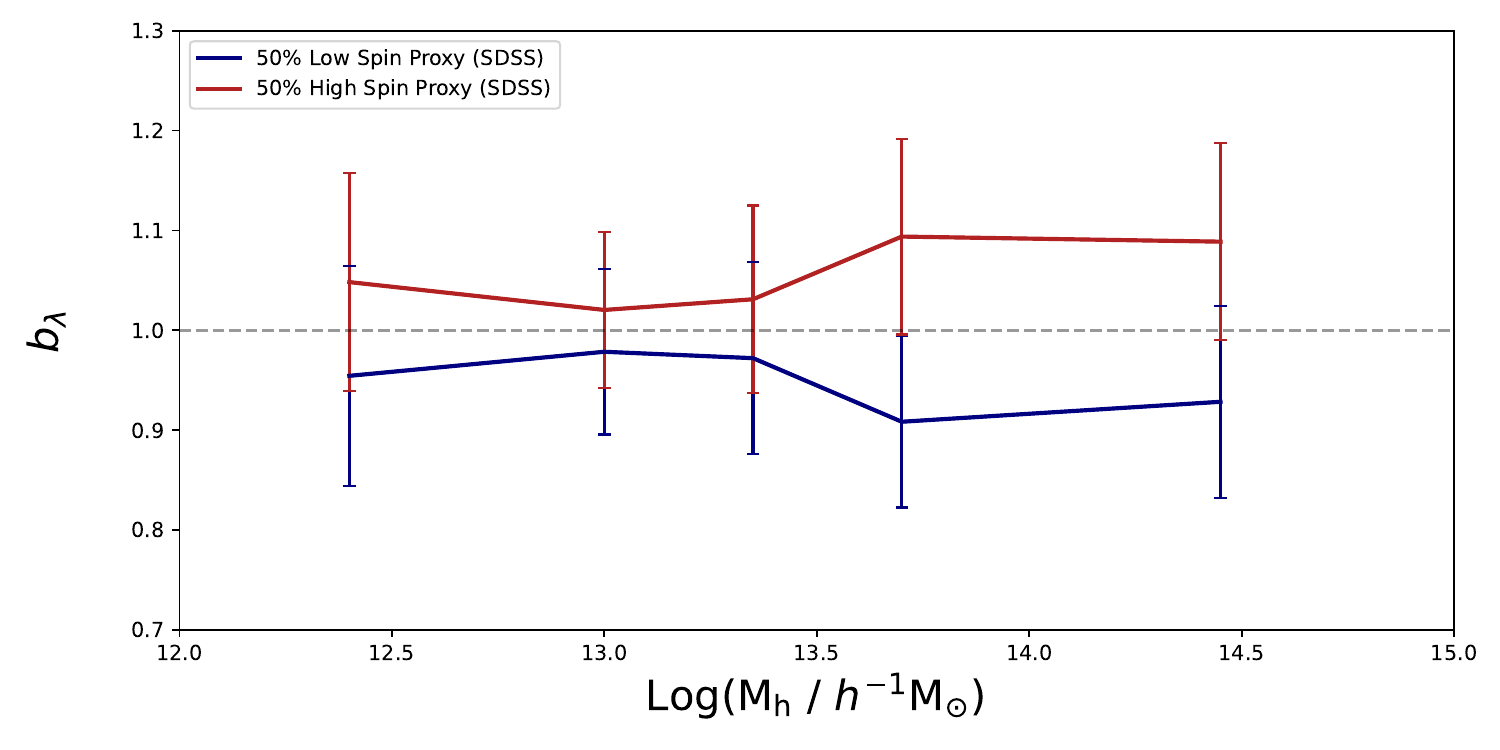}
    \caption{Spin bias signal measured from the SDSS LowZ group catalog using the halo spin proxy, which is based on the Spin-Orbit Coherence method. Red symbols and lines show results for the 50$\%$ higher spin proxy subset, while blue symbols and lines represent groups in the 50$\%$ lower spin proxy subset. Error computation details are provided in the text.} 
    \label{fig:sdssbias}
\end{figure*}

Having confirmed that the spin proxy can be used to measure spin bias in the mock, we now apply the same methodology to the observations. Figure \ref{fig:sdssbias} shows the spin bias signal measured from the SDSS LowZ group catalog, with red and blue symbols representing groups with high and low values of the spin proxy, respectively (based on 50$\%$ subsets). Our results show that the subset of groups with higher spin proxy tend to be more clustered than their lower-spin counterparts at fixed group mass, over a wide range of group masses (above $M_{\mathrm{h}} > 10^{12}$ $h^{-1}\mathrm{M}_{\odot}$). 

To assess the significance of the aforementioned measurement, we use the following procedure. First, we obtain a bias distribution by sampling the group catalog 100 times and measuring the relative bias for each of the resulting samples. Using this technique, we create two separate distributions for the high spin proxy and low spin proxy subsets. Then, to assess how often the red measurements are greater than the blue measurements, we randomly sample a red and a blue point from each distribution and record how often the red point is above the blue point, repeating this procedure 1 million times. From these 1 million red-blue pairs, we calculate the probability that the red points have  higher relative bias than the blue points. This is denoted `$P(Red > Blue)$'.

For our entire mass range, this procedure yields $P(Red > Blue) = 0.787$. For groups with masses $M_{\mathrm{h}} > 10^{13.2}$ $h^{-1}\mathrm{M}_{\odot}$, the probability rises to $P(Red > Blue) = 0.850$ (recall that this is the mass range where we expect a higher correlation between our proxy and $\lambda$, see Fig. \ref{fig:spincorr}). For individual mass bins we find $P(Red > Blue) =75.3\%, \, 59.1\%, \, 69.0\%, \, 93.7\%$ and 89.8$\%$, from low to high mass. We note that the bin widths for the SDSS measurement are larger because the sample is smaller. 

It is clear that there are some differences between results from observations and simulations. These are to be expected because of the inherent challenges of the measurement, such as fiber collisions near cluster centers and observational uncertainties in the masses of groups, to name but a few. It is encouraging, however, that the high-spin proxy measurements of relative bias are consistently above the low-spin proxy results, across the mass range for 5 independent group mass bins.

\subsection{Parameter dependencies of the observational probe} \label{sec:statresult}

\begin{figure*}[!h]
    \centering
    \includegraphics[width=85mm]{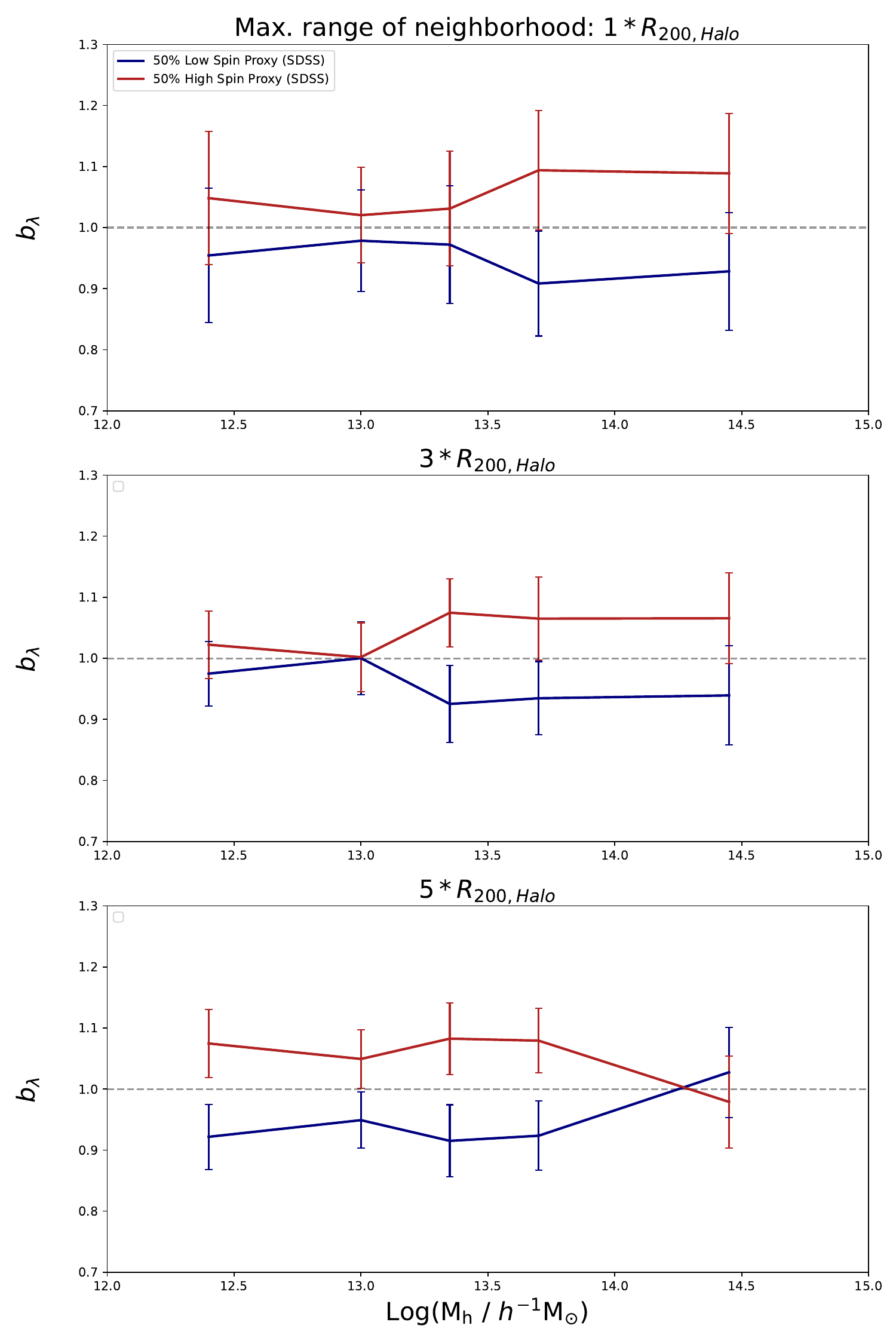}
    \includegraphics[width=85mm]{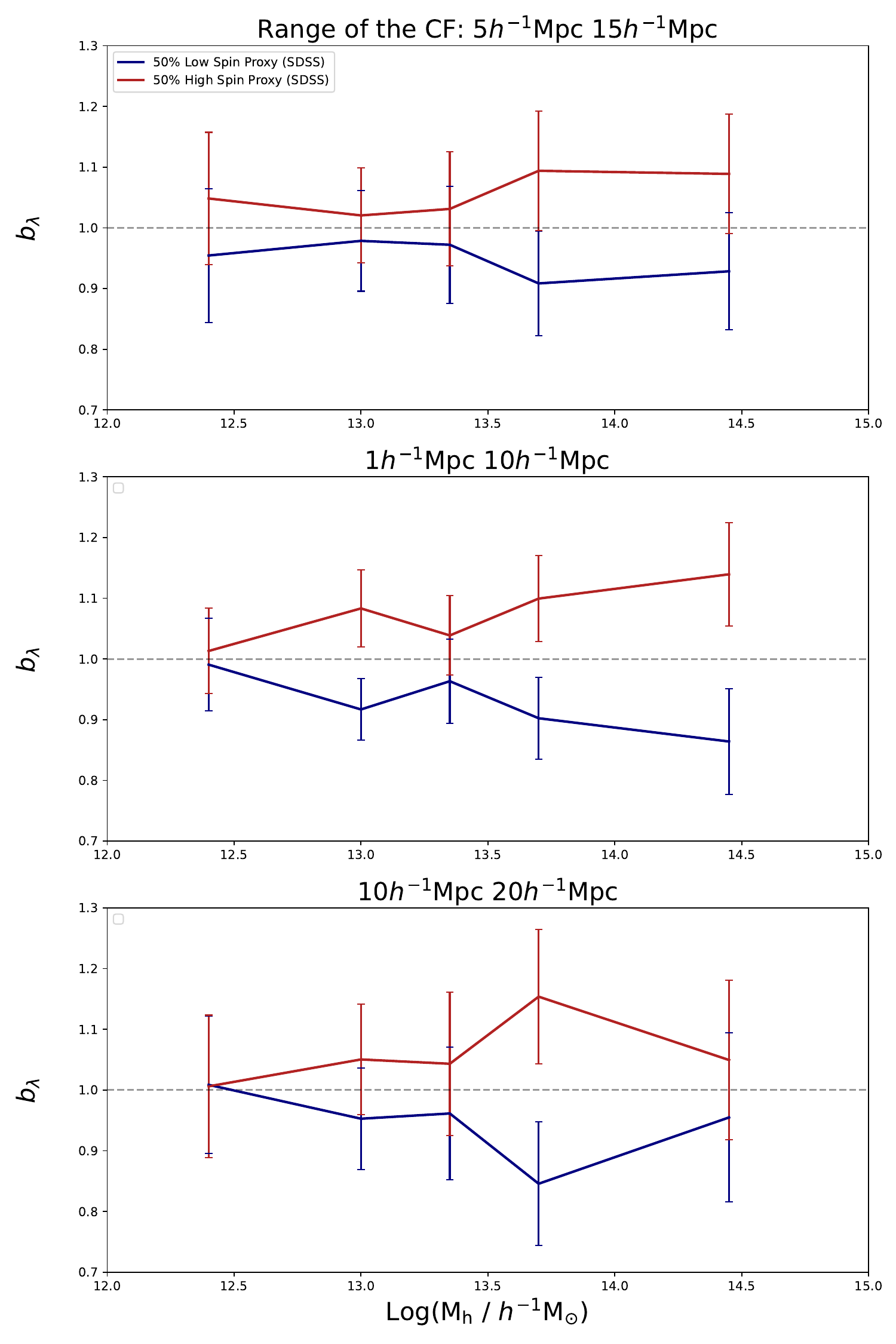}
    \caption{Similar to Fig. \ref{fig:sdssbias} but varying the configuration parameters of our spin proxy. The left column shows the impact of  varying the distance range of neighboring galaxies, while the right column displays results assuming different scale intervals for the computation of the cross-correlation function. Note that the fiducial result is shown in the top panels.}
    \label{fig:sdssbias_var}
\end{figure*}

\begin{figure*}[!h]
    \centering
    \includegraphics[width=85mm]{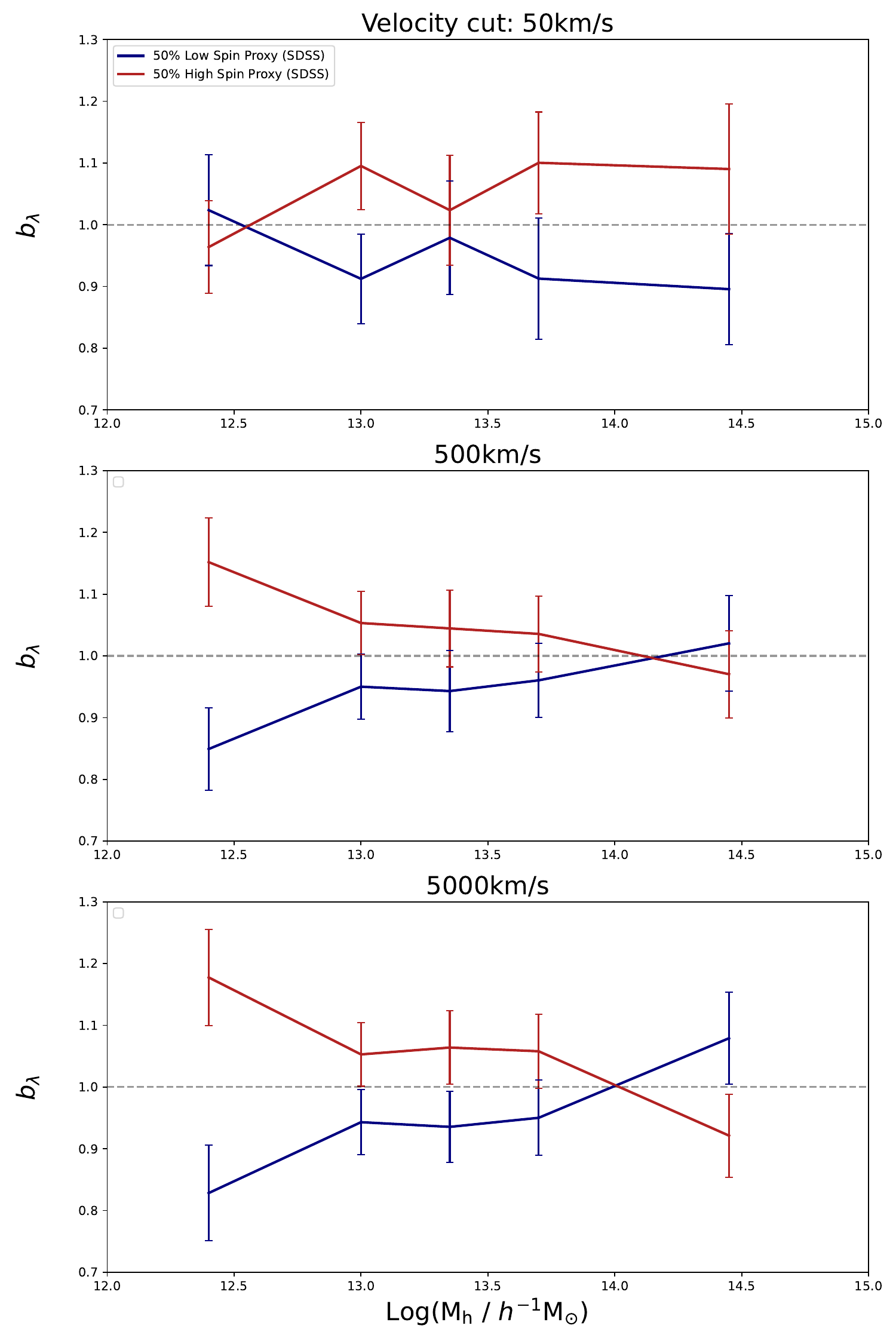}
    \includegraphics[width=85mm]{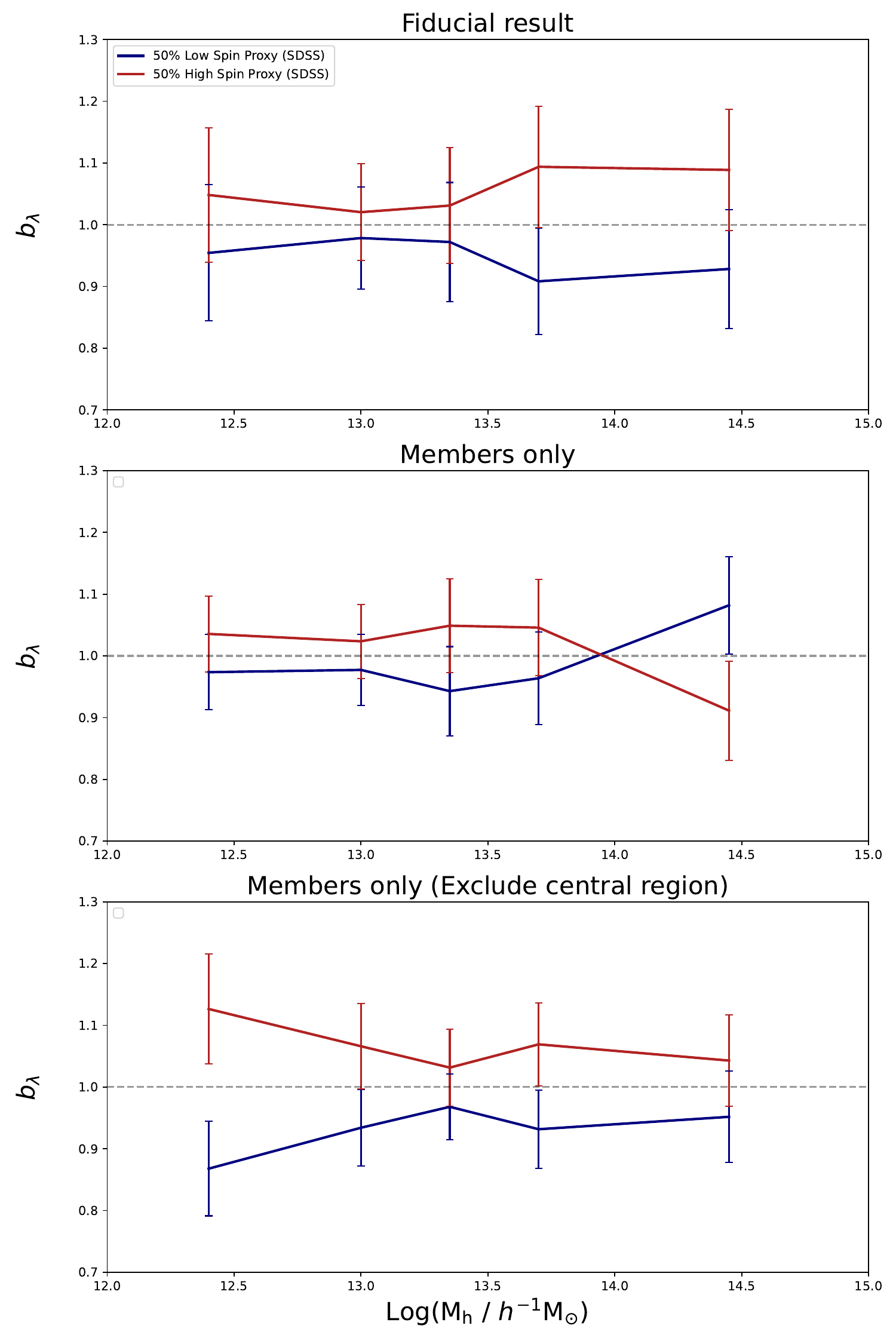}
    \caption{Similar to Fig. \ref{fig:sdssbias_var} but varying the velocity cut (left column) and our membership criteria (right column). The lower right panel shows the result of excluding galaxies within 0.3 $R_{200}$ of the groups. For comparison, the fiducial result is shown in the upper right panel.}
    \label{fig:sdssbias_appen}
\end{figure*}

In Fig. \ref{fig:sdssbias}, we chose a set of fiducial parameters to measure spin bias, which are listed in Table \ref{tab:config}. These were found to be effective when testing on our TNG300 mock. Here, we allow some of the parameters to vary to evaluate the effect on the measured spin bias and to better judge the sensitivity of the secondary bias signal to these parameters\footnote{It is important to bear in mind that, given the challenges involved and the pioneering nature of this analysis, our primary goal is to assess whether a general trend with the spin proxy exists. Reproducing the exact shape of the spin bias is beyond the scope of this paper.}. 

The results of varying the projected radius and the scales where the cross-correlation function is computed are shown in Fig. \ref{fig:sdssbias_var}. We first tried varying the range in which neighbor galaxies are identified, i.e., from $1R_{200}$, to $3R_{200}$, and up to $5R_{200}$ from the halo/group (see left column). We do not report any significant qualitative changes in the results. Next, we tried varying the range of scales over which the cross-correlation function is computed (as indicated in the figure subtitle). Once more, we do not see a significant qualitative difference between the three selections. This suggests that the measurement of spin bias based on our spin proxy is not highly sensitive to these parameter choices, at least as far as the general trend is concern. Furthermore, the high spin proxy sample is consistently and significantly above the low spin proxy sample across the majority of the mass bins. It appears, therefore, that our results are not a special case based on fine tuning of parameters.

Next, we tested the impact of varying the depth of the cylinder used to select neighbors, as well as whether including all group members -- as defined by the SDSS LowZ catalog \citep{Lim2017} -- affects the measurement. In the group catalog, group membership is determined by identifying the galaxies within the radius of $R_{200}$ of each halo, after compensating for the Fingers of God (FoG) effect. This effect makes the distribution of galaxies inside groups in redshift space to appear elongated along the line of sight. This FoG effect is primarily caused by the peculiar motion of each galaxy within the gravitational potential of the host halo. Therefore, if we select a narrow velocity cut from the halo center velocity, as done in our fiducial set up, we exclude the galaxies with the highest peculiar velocities.  However, as described in Sect. 3.1, this choice was made to concentrate on the more relaxed members of their group, which were shown to provide stronger spin proxy signal in our mock tests. This choice also mitigates environmental effects. To evaluate the effect of this velocity cut, we tested three different values; $50$,  $500$, and $5,000 \, km/s$.

An alternative approach would be to consider all members of the group as provided by the SDSS LowZ catalog \citep{Lim2017}. With this definition, there is a possibility of contamination from the surrounding environment. On the other hand, fiber collisions can produce incompleteness, especially in crowded fields, such as in cluster cores. To test the impact of fiber collisions, we conducted an additional test that excludes member galaxies within the inner $0.3\,R_{200}$ range of each halo. The results of all these velocity and membership-related tests are presented in Fig. \ref{fig:sdssbias_appen}. Despite some quantitative differences in the shapes of individual trends, the general result remain unchanged, as the higher relative bias of high spin proxy subset is visible across all five configurations. 

Additionally, we tested the impact of using group masses derived from an alternative approach, i.e., a luminosity-based method, where halo mass is traced by central galaxy luminosity instead of galaxy mass. We again obtained similar results in terms of the general trend. This uniform outcome shows that our main result is insensitive to the particular choices of the measurement. We note, again, that differences in the mass dependence of the signal are expected, given the challenges associated with this measurement.

\section{Discussion} \label{sec:dcuss}

Our results demonstrate that we are on a promising path toward an observational detection of halo spin bias using the SOC method. Upcoming surveys with larger volumes and improved statistics have the potential to confirm these preliminary findings and offer new opportunities for studying secondary bias.

While these preliminary results are encouraging, there are several aspects that can contribute to limiting the significance of our results. In observations, the more the spin vector of clusters align with the line of sight, the harder it is to estimate neighbor motion, which obviously weakens the correlation between our spin proxy and $\lambda$. Also, cluster statistics affect the correlation function, particularly because we are employing volume limited samples. We expect that these limitations can be mitigated with upcoming galaxy surveys.

Although we have only concentrated on the spin bias signal at face value, we note the possibility that other secondary parameters could also be at play, and potentially contribute to any signal that is detected by the SOC method. One example is a well known dependency of secondary bias on the local environment (e.g. \citealt{Paranjape2018, MonteroRodriguez2024}). In the future, we hope to have sufficient statistics to enable us to control for other secondary parameters, in order to cleanly investigate spin bias. Note that secondary bias connects internal halo properties with large-scale bias. So, it is important to remove any local environmental effect that could masquerade as secondary  bias signal. 

Another interesting point of discussion is that an inversion of the signal was reported in \cite{SatoPolito2019} using the MultiDark simulations, albeit at a low halo mass, i.e., $M_{\mathrm h} \simeq 10^{11.5} \, h^{-1}\mathrm{M}_{\odot}$. As we focus mainly on the higher mass range of groups in this study, in general we do not see this inversion. The physical origin of this low-mass behavior was subsequently revealed in \cite{Tucci2021}, where the authors evaluate the effect of splashback halos on the signal. These low-mass, low-spin halos that are found close to massive halos can create an inversion of the signal. The effect of splashback halos on our analysis is an interesting aspect that could potentially be addressed in a follow-up work. 

As mentioned before, it is challenging to determine if a clear dependence of spin bias on group mass exists, using the current data. Note, that in large volume simulations, it has been shown that spin bias increases towards the high mass end (i.e., the relative biases of the high and low spin subsets become progressively more separated). By refining our methods, and including additional data sets \citep{2017A&A...602A.100T, 2020A&A...636A..61R, 2024A&A...686A.157N}, we expect to be able to probe this mass dependence in the near future.

\section {Summary and conclusions} \label{sec:concl}

It is well known that the clustering of DM halos depends primarily on halo mass: more massive halos are more tightly clustered then their less massive counterparts \citep{Press1974,ShethTormen1999,Sheth2001,ShethTormen2002}. If we control for this primary effect, additional dependencies on secondary halo properties -are expected, as seen in numerical simulations (see, e.g., \citealt{Sheth2004,gao2005,Wechsler2006,Gao2007,Angulo2008,2008Li,faltenbacher2010, Lazeyras2017,2018Salcedo,han2018,Mao2018, SatoPolito2019, Johnson2019, Ramakrishnan2019,MonteroDorta2020B,Tucci2021,MonteroDorta2021_mah,MonteroRodriguez2024, MonteroDorta2025}). In this work, we present, for the first time, a methodology to measure halo spin bias in observations -- that is, the secondary dependence of halo bias on halo spin at fixed halo mass. We tested our procedure on TNG300 and applied it to the SDSS observational data.

As we cannot directly observe the true spin of groups and clusters -- the observable counterparts of halos -- we have developed a spin proxy based on what we call Spin-Orbit Coherence (SOC). This effect, previously studied in simulations \citep{YKim2022b}, refers to the correlation between the orbital motion of neighboring galaxies -- both around and within a target group -- and the magnitude and direction of the group's spin vector. 
 
To test the feasibility of using this SOC-based spin proxy in observations, we first evaluated the performance of the proxy using a mock catalog constructed from the TNG300 simulation. We confirmed that, despite the projection effects that are inherent to an observational catalog, the spin proxy displays a significant correlation with the halo spin parameter $\lambda$. This correlation is sufficient preserve the halo spin bias effect. In essence, halos with higher spin proxy at fixed halo mass have higher bias than the halo population with lower spin proxy. In TNG300, we experimented with different configurations to optimize the methodology. The strongest signal was obtained using all host halos with total masses greater than $10^{12}$ $h^{-1}$M$_\odot$, and neighbor/satellite subhalos more massive than $10^{9}$ $h^{-1}$M$_\odot$. We also only considered neighbors within $1R_{200}$ of the host halo, and the spin bias is measured from the cross-correlation function over a range of scales between 5 to 15 $h^{-1}$Mpc.

Given the promising outcome with the mocks, the next step of our analysis was to apply our methodology to an observational catalog. We used the SDSS LowZ sample \citep{Lim2017} in combination with the \citet{Yang2005} group catalog. As in TNG300, we split groups based on the spin proxy, and measure their relative bias at fixed group mass. We report indications of spin bias 
across the entire mass range considered. For the most massive groups ($\log_{10}(M_{\rm h}[h^{-1}\rm M_\odot])\gtrsim 13$), we measure a significance of 85$\%$. This value is obtained 
by randomly sampling the bias measurement of the high and low spin proxy subsets within each mass bin. In 85$\%$ of the cases, the higher spin proxy subset has a higher relative bias. 

With future data sets, this methodology could provide the first observational detection of halo spin bias. Moreover, it would be one of the first claims of secondary halo bias detection to this day (see, e.g.,\citealt{Miyatake2016,Lin2016,MonteroDorta2017B,Niemiec2018,Sunayama2022, Obuljen2020,Salcedo2022,Wang2022}, for more context). Importantly, we have verified that our results are not too sensitive to changes in the parameters that define the proxy, i.e., the 
the distance range for neighbors, the scales employed in the correlation function estimation and other choices. 

Our pioneering study opens up several new research avenues. The spin bias measurement can be potentially used as a new cosmological test. This research line could be strengthened if the same measurement is performed in other spectroscopic data sets at higher redshift, so that the evolution of the spin-bias signal can be determined. Furthermore, measuring halo spin for a sizable sample of clusters and groups would allow us to study how these systems gain spin and how this process depends on the cosmic web. Our long term goal is to use this work as a foundation for future endeavors to bridge the gap between internal halo properties and their surrounding environments, over diverse scales. 

\begin{acknowledgements}

Yigon Kim has supported by National Research Foundation of Korea (NRF), through grants funded by the Korean government (MSIT) (Nos. 2022R1A4A3031306), during this work. Yigon also thanks Kyungpook National University SPHEREx Basic Research Lab and Seoul National University ExgalCos group for their kind support. Antonio D. Montero-Dorta thanks Fondecyt for financial support through the Fondecyt Regular 2021 grant 1210612. Rory Smith acknowledges financial support from FONDECYT Regular 2023 project No. 1230441 and also gratefully acknowledges financial support from ANID - MILENIO NCN2024$\_$112.

\end{acknowledgements}

\bibliographystyle{aa} 
\bibliography{main.bib}

\begin{thebibliography}{70}
\expandafter\ifx\csname natexlab\endcsname\relax\def\natexlab#1{#1}\fi

\bibitem[{{Albareti} {et~al.}(2017){Albareti}, {Allende Prieto}, {Almeida}, {Anders}, {Anderson}, {Andrews}, {Arag{\'o}n-Salamanca}, {Argudo-Fern{\'a}ndez}, {Armengaud}, {Aubourg}, {Avila-Reese}, {Badenes}, {Bailey}, {Barbuy}, {Barger}, {Barrera-Ballesteros}, {Bartosz}, {Basu}, {Bates}, {Battaglia}, {Baumgarten}, {Baur}, {Bautista}, {Beers}, {Belfiore}, {Bershady}, {Bertran de Lis}, {Bird}, {Bizyaev}, {Blanc}, {Blanton}, {Blomqvist}, {Bolton}, {Borissova}, {Bovy}, {Brandt}, {Brinkmann}, {Brownstein}, {Bundy}, {Burtin}, {Busca}, {Camacho Chavez}, {Cano D{\'\i}az}, {Cappellari}, {Carrera}, {Chen}, {Cherinka}, {Cheung}, {Chiappini}, {Chojnowski}, {Chuang}, {Chung}, {Cirolini}, {Clerc}, {Cohen}, {Comerford}, {Comparat}, {Correa do Nascimento}, {Cousinou}, {Covey}, {Crane}, {Croft}, {Cunha}, {Darling}, {Davidson}, {Dawson}, {Da Costa}, {Da Silva Ilha}, {Deconto Machado}, {Delubac}, {De Lee}, {De la Macorra}, {De la Torre}, {Diamond-Stanic}, {Donor}, {Downes}, {Drory}, {Du}, {Du Mas des Bourboux}, {Dwelly},
  {Ebelke}, {Eigenbrot}, {Eisenstein}, {Elsworth}, {Emsellem}, {Eracleous}, {Escoffier}, {Evans}, {Falc{\'o}n-Barroso}, {Fan}, {Favole}, {Fernandez-Alvar}, {Fernandez-Trincado}, {Feuillet}, {Fleming}, {Font-Ribera}, {Freischlad}, {Frinchaboy}, {Fu}, {Gao}, {Garcia}, {Garcia-Dias}, {Garcia-Hern{\'a}ndez}, {Garcia P{\'e}rez}, {Gaulme}, {Ge}, {Geisler}, {Gillespie}, {Gil Marin}, {Girardi}, {Goddard}, {Gomez Maqueo Chew}, {Gonzalez-Perez}, {Grabowski}, {Green}, {Grier}, {Grier}, {Guo}, {Guy}, {Hagen}, {Hall}, {Harding}, {Harley}, {Hasselquist}, {Hawley}, {Hayes}, {Hearty}, {Hekker}, {Hernandez Toledo}, {Ho}, {Hogg}, {Holley-Bockelmann}, {Holtzman}, {Holzer}, {Hu}, {Huber}, {Hutchinson}, {Hwang}, {Ibarra-Medel}, {Ivans}, {Ivory}, {Jaehnig}, {Jensen}, {Johnson}, {Jones}, {Jullo}, {Kallinger}, {Kinemuchi}, {Kirkby}, {Klaene}, {Kneib}, {Kollmeier}, {Lacerna}, {Lane}, {Lang}, {Laurent}, {Law}, {Leauthaud}, {Le Goff}, {Li}, {Li}, {Li}, {Li}, {Liang}, {Liang}, {Lima}, {Lin}, {Lin}, {Lin}, {Liu}, {Long}, {Lucatello},
  {MacDonald}, {MacLeod}, {Mackereth}, {Mahadevan}, {Maia}, {Maiolino}, {Majewski}, {Malanushenko}, {Malanushenko}, {Mallmann}, {Manchado}, {Maraston}, {Marques-Chaves}, {Martinez Valpuesta}, {Masters}, {Mathur}, {McGreer}, {Merloni}, {Merrifield}, {M{\'e}sz{\'a}ros}, {Meza}, {Miglio}, {Minchev}, {Molaverdikhani}, {Montero-Dorta}, {Mosser}, {Muna}, {Myers}, {Nair}, {Nandra}, {Ness}, {Newman}, {Nichol}, {Nidever}, {Nitschelm}, {O'Connell}, {Oravetz}, {Oravetz}, {Pace}, {Padilla}, {Palanque-Delabrouille}, {Pan}, {Parejko}, {Paris}, {Park}, {Peacock}, {Peirani}, {Pellejero-Ibanez}, {Penny}, {Percival}, {Percival}, {Perez-Fournon}, {Petitjean}, {Pieri}, {Pinsonneault}, {Pisani}, {Prada}, {Prakash}, {Price-Jones}, {Raddick}, {Rahman}, {Raichoor}, {Barboza Rembold}, {Reyna}, {Rich}, {Richstein}, {Ridl}, {Riffel}, {Riffel}, {Rix}, {Robin}, {Rockosi}, {Rodr{\'\i}guez-Torres}, {Rodrigues}, {Roe}, {Roman Lopes}, {Rom{\'a}n-Z{\'u}{\~n}iga}, {Ross}, {Rossi}, {Ruan}, {Ruggeri}, {Runnoe}, {Salazar-Albornoz}, {Salvato},
  {Sanchez}, {Sanchez}, {Sanchez-Gallego}, {Santiago}, {Schiavon}, {Schimoia}, {Schlafly}, {Schlegel}, {Schneider}, {Sch{\"o}nrich}, {Schultheis}, {Schwope}, {Seo}, {Serenelli}, {Sesar}, {Shao}, {Shetrone}, {Shull}, {Silva Aguirre}, {Skrutskie}, {Slosar}, {Smith}, {Smith}, {Sobeck}, {Somers}, {Souto}, {Stark}, {Stassun}, {Steinmetz}, {Stello}, {Storchi Bergmann}, {Strauss}, {Streblyanska}, {Stringfellow}, {Suarez}, {Sun}, {Taghizadeh-Popp}, {Tang}, {Tao}, {Tayar}, {Tembe}, {Thomas}, {Tinker}, {Tojeiro}, {Tremonti}, {Troup}, {Trump}, {Unda-Sanzana}, {Valenzuela}, {Van den Bosch}, {Vargas-Maga{\~n}a}, {Vazquez}, {Villanova}, {Vivek}, {Vogt}, {Wake}, {Walterbos}, {Wang}, {Wang}, {Weaver}, {Weijmans}, {Weinberg}, {Westfall}, {Whelan}, {Wilcots}, {Wild}, {Williams}, {Wilson}, {Wood-Vasey}, {Wylezalek}, {Xiao}, {Yan}, {Yang}, {Ybarra}, {Yeche}, {Yuan}, {Zakamska}, {Zamora}, {Zasowski}, {Zhang}, {Zhao}, {Zhao}, {Zheng}, {Zheng}, {Zhou}, {Zhu}, {Zinn}, \& {Zou}}]{SDSSLowZ}
{Albareti}, F.~D., {Allende Prieto}, C., {Almeida}, A., {et~al.} 2017, \apjs, 233, 25

\bibitem[{{Angulo} {et~al.}(2008){Angulo}, {Baugh}, \& {Lacey}}]{Angulo2008}
{Angulo}, R.~E., {Baugh}, C.~M., \& {Lacey}, C.~G. 2008, \mnras, 387, 921

\bibitem[{{Barnes} \& {Efstathiou}(1987)}]{BarnesEfstathiou1987}
{Barnes}, J. \& {Efstathiou}, G. 1987, \apj, 319, 575

\bibitem[{{Bell} {et~al.}(2003){Bell}, {McIntosh}, {Katz}, \& {Weinberg}}]{Bell2003}
{Bell}, E.~F., {McIntosh}, D.~H., {Katz}, N., \& {Weinberg}, M.~D. 2003, \apjs, 149, 289

\bibitem[{{Bullock} {et~al.}(2001){Bullock}, {Dekel}, {Kolatt}, {Kravtsov}, {Klypin}, {Porciani}, \& {Primack}}]{Bullock2001}
{Bullock}, J.~S., {Dekel}, A., {Kolatt}, T.~S., {et~al.} 2001, \apj, 555, 240

\bibitem[{{Dalal} {et~al.}(2008){Dalal}, {White}, {Bond}, \& {Shirokov}}]{Dalal2008}
{Dalal}, N., {White}, M., {Bond}, J.~R., \& {Shirokov}, A. 2008, \apj, 687, 12

\bibitem[{{Faltenbacher} \& {White}(2010)}]{faltenbacher2010}
{Faltenbacher}, A. \& {White}, S. D.~M. 2010, \apj, 708, 469

\bibitem[{{Gao} {et~al.}(2005){Gao}, {Springel}, \& {White}}]{gao2005}
{Gao}, L., {Springel}, V., \& {White}, S.~D.~M. 2005, \mnras, 363, L66

\bibitem[{{Gao} \& {White}(2007)}]{Gao2007}
{Gao}, L. \& {White}, S.~D.~M. 2007, \mnras, 377, L5

\bibitem[{{Genel} {et~al.}(2014){Genel}, {Vogelsberger}, {Springel}, {Sijacki}, {Nelson}, {Snyder}, {Rodriguez-Gomez}, {Torrey}, \& {Hernquist}}]{Genel2014}
{Genel}, S., {Vogelsberger}, M., {Springel}, V., {et~al.} 2014, \mnras, 445, 175

\bibitem[{{Han} {et~al.}(2019){Han}, {Li}, {Jing}, {Nishimichi}, {Wang}, \& {Jiang}}]{han2018}
{Han}, J., {Li}, Y., {Jing}, Y., {et~al.} 2019, \mnras, 482, 1900

\bibitem[{{Johnson} {et~al.}(2019){Johnson}, {Maller}, {Berlind}, {Sinha}, \& {Holley-Bockelmann}}]{Johnson2019}
{Johnson}, J.~W., {Maller}, A.~H., {Berlind}, A.~A., {Sinha}, M., \& {Holley-Bockelmann}, J.~K. 2019, \mnras, 486, 1156

\bibitem[{{Kim} {et~al.}(2022){Kim}, {Smith}, \& {Shin}}]{YKim2022b}
{Kim}, Y., {Smith}, R., \& {Shin}, J. 2022, \apj, 935, 71

\bibitem[{{Landy} \& {Szalay}(1993)}]{1993ApJ...412...64L}
{Landy}, S.~D. \& {Szalay}, A.~S. 1993, \apj, 412, 64

\bibitem[{{Lazeyras} {et~al.}(2017){Lazeyras}, {Musso}, \& {Schmidt}}]{Lazeyras2017}
{Lazeyras}, T., {Musso}, M., \& {Schmidt}, F. 2017, \jcap, 2017, 059

\bibitem[{{Lee} {et~al.}(2019{\natexlab{a}}){Lee}, {Pak}, {Lee}, \& {Song}}]{JLee2019a}
{Lee}, J.~H., {Pak}, M., {Lee}, H.-R., \& {Song}, H. 2019{\natexlab{a}}, \apj, 872, 78

\bibitem[{{Lee} {et~al.}(2019{\natexlab{b}}){Lee}, {Pak}, {Song}, {Lee}, {Kim}, \& {Jeong}}]{JLee2019b}
{Lee}, J.~H., {Pak}, M., {Song}, H., {et~al.} 2019{\natexlab{b}}, \apj, 884, 104

\bibitem[{{Li} {et~al.}(2008){Li}, {Mo}, \& {Gao}}]{2008Li}
{Li}, Y., {Mo}, H.~J., \& {Gao}, L. 2008, \mnras, 389, 1419

\bibitem[{{Lim} {et~al.}(2017){Lim}, {Mo}, {Lu}, {Wang}, \& {Yang}}]{Lim2017}
{Lim}, S.~H., {Mo}, H.~J., {Lu}, Y., {Wang}, H., \& {Yang}, X. 2017, \mnras, 470, 2982

\bibitem[{{Lin} {et~al.}(2016){Lin}, {Mandelbaum}, {Huang}, {Huang}, {Dalal}, {Diemer}, {Jian}, \& {Kravtsov}}]{Lin2016}
{Lin}, Y.-T., {Mandelbaum}, R., {Huang}, Y.-H., {et~al.} 2016, \apj, 819, 119

\bibitem[{{Mao} {et~al.}(2018){Mao}, {Zentner}, \& {Wechsler}}]{Mao2018}
{Mao}, Y.-Y., {Zentner}, A.~R., \& {Wechsler}, R.~H. 2018, \mnras, 474, 5143

\bibitem[{{Marinacci} {et~al.}(2018){Marinacci}, {Vogelsberger}, {Pakmor}, {Torrey}, {Springel}, {Hernquist}, {Nelson}, {Weinberger}, {Pillepich}, {Naiman}, \& {Genel}}]{Marinacci2018}
{Marinacci}, F., {Vogelsberger}, M., {Pakmor}, R., {et~al.} 2018, \mnras, 480, 5113

\bibitem[{{Miyatake} {et~al.}(2016){Miyatake}, {More}, {Takada}, {Spergel}, {Mandelbaum}, {Rykoff}, \& {Rozo}}]{Miyatake2016}
{Miyatake}, H., {More}, S., {Takada}, M., {et~al.} 2016, \prl, 116, 041301

\bibitem[{{Mo} \& {White}(1996)}]{Mo1996}
{Mo}, H.~J. \& {White}, S.~D.~M. 1996, \mnras, 282, 347

\bibitem[{{Montero-Dorta}(2021)}]{MonteroDorta2021_hsb}
{Montero-Dorta}, A.~D. 2021, Boletin de la Asociacion Argentina de Astronomia La Plata Argentina, 62, 159

\bibitem[{{Montero-Dorta} {et~al.}(2021{\natexlab{a}}){Montero-Dorta}, {Artale}, {Abramo}, \& {Tucci}}]{MonteroDorta2021_SZ}
{Montero-Dorta}, A.~D., {Artale}, M.~C., {Abramo}, L.~R., \& {Tucci}, B. 2021{\natexlab{a}}, \mnras, 504, 4568

\bibitem[{{Montero-Dorta} {et~al.}(2020){Montero-Dorta}, {Artale}, {Abramo}, {Tucci}, {Padilla}, {Sato-Polito}, {Lacerna}, {Rodriguez}, \& {Angulo}}]{MonteroDorta2020B}
{Montero-Dorta}, A.~D., {Artale}, M.~C., {Abramo}, L.~R., {et~al.} 2020, \mnras, 496, 1182

\bibitem[{{Montero-Dorta} {et~al.}(2021{\natexlab{b}}){Montero-Dorta}, {Chaves-Montero}, {Artale}, \& {Favole}}]{MonteroDorta2021_mah}
{Montero-Dorta}, A.~D., {Chaves-Montero}, J., {Artale}, M.~C., \& {Favole}, G. 2021{\natexlab{b}}, \mnras, 508, 940

\bibitem[{{Montero-Dorta} {et~al.}(2025){Montero-Dorta}, {Contreras}, {Celeste Artale}, {Rodriguez}, \& {Favole}}]{MonteroDorta2025}
{Montero-Dorta}, A.~D., {Contreras}, S., {Celeste Artale}, M., {Rodriguez}, F., \& {Favole}, G. 2025, \aap, 695, A159

\bibitem[{{Montero-Dorta} {et~al.}(2017){Montero-Dorta}, {P{\'e}rez}, {Prada}, {Rodr{\'{\i}}guez-Torres}, {Favole}, {Klypin}, {Cid Fernandes}, {Gonz{\'a}lez Delgado}, {Dom{\'{\i}}nguez}, {Bolton}, {Garc{\'{\i}}a-Benito}, {Jullo}, \& {Niemiec}}]{MonteroDorta2017B}
{Montero-Dorta}, A.~D., {P{\'e}rez}, E., {Prada}, F., {et~al.} 2017, \apjl, 848, L2

\bibitem[{{Montero-Dorta} \& {Rodriguez}(2024)}]{MonteroRodriguez2024}
{Montero-Dorta}, A.~D. \& {Rodriguez}, F. 2024, \mnras, 531, 290

\bibitem[{{Mroczkowski} {et~al.}(2019){Mroczkowski}, {Nagai}, {Basu}, {Chluba}, {Sayers}, {Adam}, {Churazov}, {Crites}, {Di Mascolo}, {Eckert}, {Macias-Perez}, {Mayet}, {Perotto}, {Pointecouteau}, {Romero}, {Ruppin}, {Scannapieco}, \& {ZuHone}}]{Mroczkowski2019}
{Mroczkowski}, T., {Nagai}, D., {Basu}, K., {et~al.} 2019, \ssr, 215, 17

\bibitem[{{Naiman} {et~al.}(2018){Naiman}, {Pillepich}, {Springel}, {Ramirez-Ruiz}, {Torrey}, {Vogelsberger}, {Pakmor}, {Nelson}, {Marinacci}, {Hernquist}, {Weinberger}, \& {Genel}}]{Naiman2018}
{Naiman}, J.~P., {Pillepich}, A., {Springel}, V., {et~al.} 2018, \mnras, 477, 1206

\bibitem[{{Nelson} {et~al.}(2024){Nelson}, {Pillepich}, {Ayromlou}, {Lee}, {Lehle}, {Rohr}, \& {Truong}}]{2024A&A...686A.157N}
{Nelson}, D., {Pillepich}, A., {Ayromlou}, M., {et~al.} 2024, \aap, 686, A157

\bibitem[{{Nelson} {et~al.}(2018){Nelson}, {Pillepich}, {Springel}, {Weinberger}, {Hernquist}, {Pakmor}, {Genel}, {Torrey}, {Vogelsberger}, {Kauffmann}, {Marinacci}, \& {Naiman}}]{Nelson2018_ColorBim}
{Nelson}, D., {Pillepich}, A., {Springel}, V., {et~al.} 2018, \mnras, 475, 624

\bibitem[{{Nelson} {et~al.}(2019){Nelson}, {Springel}, {Pillepich}, {Rodriguez-Gomez}, {Torrey}, {Genel}, {Vogelsberger}, {Pakmor}, {Marinacci}, {Weinberger}, {Kelley}, {Lovell}, {Diemer}, \& {Hernquist}}]{Nelson2019}
{Nelson}, D., {Springel}, V., {Pillepich}, A., {et~al.} 2019, Computational Astrophysics and Cosmology, 6, 2

\bibitem[{{Niemiec} {et~al.}(2018){Niemiec}, {Jullo}, {Montero-Dorta}, {Prada}, {Rodriguez-Torres}, {Perez}, {Klypin}, {Erben}, {Makler}, {Moraes}, {Pereira}, \& {Shan}}]{Niemiec2018}
{Niemiec}, A., {Jullo}, E., {Montero-Dorta}, A.~D., {et~al.} 2018, \mnras [\eprint[arXiv]{1801.06551}]

\bibitem[{{Obuljen} {et~al.}(2020){Obuljen}, {Percival}, \& {Dalal}}]{Obuljen2020}
{Obuljen}, A., {Percival}, W.~J., \& {Dalal}, N. 2020, \jcap, 2020, 058

\bibitem[{{Paranjape} {et~al.}(2018){Paranjape}, {Hahn}, \& {Sheth}}]{Paranjape2018}
{Paranjape}, A., {Hahn}, O., \& {Sheth}, R.~K. 2018, \mnras, 476, 3631

\bibitem[{{Pasquali} {et~al.}(2019){Pasquali}, {Smith}, {Gallazzi}, {De Lucia}, {Zibetti}, {Hirschmann}, \& {Yi}}]{Pasquali2019}
{Pasquali}, A., {Smith}, R., {Gallazzi}, A., {et~al.} 2019, \mnras, 484, 1702

\bibitem[{{Peebles}(1969)}]{Peebles1969}
{Peebles}, P.~J.~E. 1969, \apj, 155, 393

\bibitem[{{Pillepich} {et~al.}(2018{\natexlab{a}}){Pillepich}, {Nelson}, {Hernquist}, {Springel}, {Pakmor}, {Torrey}, {Weinberger}, {Genel}, {Naiman}, {Marinacci}, \& {Vogelsberger}}]{Pillepich2018b}
{Pillepich}, A., {Nelson}, D., {Hernquist}, L., {et~al.} 2018{\natexlab{a}}, \mnras, 475, 648

\bibitem[{{Pillepich} {et~al.}(2018{\natexlab{b}}){Pillepich}, {Springel}, {Nelson}, {Genel}, {Naiman}, {Pakmor}, {Hernquist}, {Torrey}, {Vogelsberger}, {Weinberger}, \& {Marinacci}}]{Pillepich2018}
{Pillepich}, A., {Springel}, V., {Nelson}, D., {et~al.} 2018{\natexlab{b}}, \mnras, 473, 4077

\bibitem[{{Planck Collaboration} {et~al.}(2016){Planck Collaboration}, {Ade}, {Aghanim}, {Arnaud}, {Ashdown}, {Aumont}, {Baccigalupi}, {Banday}, {Barreiro}, {Bartlett}, {Bartolo}, {Battaner}, {Battye}, {Benabed}, {Beno{\^\i}t}, {Benoit-L{\'e}vy}, {Bernard}, {Bersanelli}, {Bielewicz}, {Bock}, {Bonaldi}, {Bonavera}, {Bond}, {Borrill}, {Bouchet}, {Boulanger}, {Bucher}, {Burigana}, {Butler}, {Calabrese}, {Cardoso}, {Catalano}, {Challinor}, {Chamballu}, {Chary}, {Chiang}, {Chluba}, {Christensen}, {Church}, {Clements}, {Colombi}, {Colombo}, {Combet}, {Coulais}, {Crill}, {Curto}, {Cuttaia}, {Danese}, {Davies}, {Davis}, {de Bernardis}, {de Rosa}, {de Zotti}, {Delabrouille}, {D{\'e}sert}, {Di Valentino}, {Dickinson}, {Diego}, {Dolag}, {Dole}, {Donzelli}, {Dor{\'e}}, {Douspis}, {Ducout}, {Dunkley}, {Dupac}, {Efstathiou}, {Elsner}, {En{\ss}lin}, {Eriksen}, {Farhang}, {Fergusson}, {Finelli}, {Forni}, {Frailis}, {Fraisse}, {Franceschi}, {Frejsel}, {Galeotta}, {Galli}, {Ganga}, {Gauthier}, {Gerbino}, {Ghosh}, {Giard},
  {Giraud-H{\'e}raud}, {Giusarma}, {Gjerl{\o}w}, {Gonz{\'a}lez-Nuevo}, {G{\'o}rski}, {Gratton}, {Gregorio}, {Gruppuso}, {Gudmundsson}, {Hamann}, {Hansen}, {Hanson}, {Harrison}, {Helou}, {Henrot-Versill{\'e}}, {Hern{\'a}ndez-Monteagudo}, {Herranz}, {Hildebrandt}, {Hivon}, {Hobson}, {Holmes}, {Hornstrup}, {Hovest}, {Huang}, {Huffenberger}, {Hurier}, {Jaffe}, {Jaffe}, {Jones}, {Juvela}, {Keih{\"a}nen}, {Keskitalo}, {Kisner}, {Kneissl}, {Knoche}, {Knox}, {Kunz}, {Kurki-Suonio}, {Lagache}, {L{\"a}hteenm{\"a}ki}, {Lamarre}, {Lasenby}, {Lattanzi}, {Lawrence}, {Leahy}, {Leonardi}, {Lesgourgues}, {Levrier}, {Lewis}, {Liguori}, {Lilje}, {Linden-V{\o}rnle}, {L{\'o}pez-Caniego}, {Lubin}, {Mac{\'\i}as-P{\'e}rez}, {Maggio}, {Maino}, {Mandolesi}, {Mangilli}, {Marchini}, {Maris}, {Martin}, {Martinelli}, {Mart{\'\i}nez-Gonz{\'a}lez}, {Masi}, {Matarrese}, {McGehee}, {Meinhold}, {Melchiorri}, {Melin}, {Mendes}, {Mennella}, {Migliaccio}, {Millea}, {Mitra}, {Miville-Desch{\^e}nes}, {Moneti}, {Montier}, {Morgante}, {Mortlock},
  {Moss}, {Munshi}, {Murphy}, {Naselsky}, {Nati}, {Natoli}, {Netterfield}, {N{\o}rgaard-Nielsen}, {Noviello}, {Novikov}, {Novikov}, {Oxborrow}, {Paci}, {Pagano}, {Pajot}, {Paladini}, {Paoletti}, {Partridge}, {Pasian}, {Patanchon}, {Pearson}, {Perdereau}, {Perotto}, {Perrotta}, {Pettorino}, {Piacentini}, {Piat}, {Pierpaoli}, {Pietrobon}, {Plaszczynski}, {Pointecouteau}, {Polenta}, {Popa}, {Pratt}, {Pr{\'e}zeau}, {Prunet}, {Puget}, {Rachen}, {Reach}, {Rebolo}, {Reinecke}, {Remazeilles}, {Renault}, {Renzi}, {Ristorcelli}, {Rocha}, {Rosset}, {Rossetti}, {Roudier}, {Rouill{\'e} d'Orfeuil}, {Rowan-Robinson}, {Rubi{\~n}o-Mart{\'\i}n}, {Rusholme}, {Said}, {Salvatelli}, {Salvati}, {Sandri}, {Santos}, {Savelainen}, {Savini}, {Scott}, {Seiffert}, {Serra}, {Shellard}, {Spencer}, {Spinelli}, {Stolyarov}, {Stompor}, {Sudiwala}, {Sunyaev}, {Sutton}, {Suur-Uski}, {Sygnet}, {Tauber}, {Terenzi}, {Toffolatti}, {Tomasi}, {Tristram}, {Trombetti}, {Tucci}, {Tuovinen}, {T{\"u}rler}, {Umana}, {Valenziano}, {Valiviita}, {Van Tent},
  {Vielva}, {Villa}, {Wade}, {Wandelt}, {Wehus}, {White}, {White}, {Wilkinson}, {Yvon}, {Zacchei}, \& {Zonca}}]{planck2016}
{Planck Collaboration}, {Ade}, P.~A.~R., {Aghanim}, N., {et~al.} 2016, \aap, 594, A13

\bibitem[{{Press} \& {Schechter}(1974)}]{Press1974}
{Press}, W.~H. \& {Schechter}, P. 1974, \apj, 187, 425

\bibitem[{{Ramakrishnan} {et~al.}(2019){Ramakrishnan}, {Paranjape}, {Hahn}, \& {Sheth}}]{Ramakrishnan2019}
{Ramakrishnan}, S., {Paranjape}, A., {Hahn}, O., \& {Sheth}, R.~K. 2019, \mnras, 489, 2977

\bibitem[{{Rodriguez} \& {Merch{\'a}n}(2020)}]{2020A&A...636A..61R}
{Rodriguez}, F. \& {Merch{\'a}n}, M. 2020, \aap, 636, A61

\bibitem[{{Salcedo} {et~al.}(2018){Salcedo}, {Maller}, {Berlind}, {Sinha}, {McBride}, {Behroozi}, {Wechsler}, \& {Weinberg}}]{2018Salcedo}
{Salcedo}, A.~N., {Maller}, A.~H., {Berlind}, A.~A., {et~al.} 2018, \mnras, 475, 4411

\bibitem[{{Salcedo} {et~al.}(2022){Salcedo}, {Zu}, {Zhang}, {Wang}, {Yang}, {Wu}, {Jing}, {Mo}, \& {Weinberg}}]{Salcedo2022}
{Salcedo}, A.~N., {Zu}, Y., {Zhang}, Y., {et~al.} 2022, Science China Physics, Mechanics, and Astronomy, 65, 109811

\bibitem[{{S{\'a}nchez} {et~al.}(2016){S{\'a}nchez}, {Garc{\'\i}a-Benito}, {Zibetti}, {Walcher}, {Husemann}, {Mendoza}, {Galbany}, {Falc{\'o}n-Barroso}, {Mast}, {Aceituno}, {Aguerri}, {Alves}, {Amorim}, {Ascasibar}, {Barrado-Navascues}, {Barrera-Ballesteros}, {Bekerait{\`e}}, {Bland-Hawthorn}, {Cano D{\'\i}az}, {Cid Fernandes}, {Cavichia}, {Cortijo}, {Dannerbauer}, {Demleitner}, {D{\'\i}az}, {Dettmar}, {de Lorenzo-C{\'a}ceres}, {del Olmo}, {Galazzi}, {Garc{\'\i}a-Lorenzo}, {Gil de Paz}, {Gonz{\'a}lez Delgado}, {Holmes}, {Igl{\'e}sias-P{\'a}ramo}, {Kehrig}, {Kelz}, {Kennicutt}, {Kleemann}, {Lacerda}, {L{\'o}pez Fern{\'a}ndez}, {L{\'o}pez S{\'a}nchez}, {Lyubenova}, {Marino}, {M{\'a}rquez}, {Mendez-Abreu}, {Moll{\'a}}, {Monreal-Ibero}, {Ortega Minakata}, {Torres-Papaqui}, {P{\'e}rez}, {Rosales-Ortega}, {Roth}, {S{\'a}nchez-Bl{\'a}zquez}, {Schilling}, {Spekkens}, {Vale Asari}, {van den Bosch}, {van de Ven}, {Vilchez}, {Wild}, {Wisotzki}, {Y{\i}ld{\i}r{\i}m}, \& {Ziegler}}]{califa2}
{S{\'a}nchez}, S.~F., {Garc{\'\i}a-Benito}, R., {Zibetti}, S., {et~al.} 2016, \aap, 594, A36

\bibitem[{{S{\'a}nchez} {et~al.}(2012){S{\'a}nchez}, {Kennicutt}, {Gil de Paz}, {van de Ven}, {V{\'\i}lchez}, {Wisotzki}, {Walcher}, {Mast}, {Aguerri}, {Albiol-P{\'e}rez}, {Alonso-Herrero}, {Alves}, {Bakos}, {Bart{\'a}kov{\'a}}, {Bland-Hawthorn}, {Boselli}, {Bomans}, {Castillo-Morales}, {Cortijo-Ferrero}, {de Lorenzo-C{\'a}ceres}, {Del Olmo}, {Dettmar}, {D{\'\i}az}, {Ellis}, {Falc{\'o}n-Barroso}, {Flores}, {Gallazzi}, {Garc{\'\i}a-Lorenzo}, {Gonz{\'a}lez Delgado}, {Gruel}, {Haines}, {Hao}, {Husemann}, {Igl{\'e}sias-P{\'a}ramo}, {Jahnke}, {Johnson}, {Jungwiert}, {Kalinova}, {Kehrig}, {Kupko}, {L{\'o}pez-S{\'a}nchez}, {Lyubenova}, {Marino}, {M{\'a}rmol-Queralt{\'o}}, {M{\'a}rquez}, {Masegosa}, {Meidt}, {Mendez-Abreu}, {Monreal-Ibero}, {Montijo}, {Mour{\~a}o}, {Palacios-Navarro}, {Papaderos}, {Pasquali}, {Peletier}, {P{\'e}rez}, {P{\'e}rez}, {Quirrenbach}, {Rela{\~n}o}, {Rosales-Ortega}, {Roth}, {Ruiz-Lara}, {S{\'a}nchez-Bl{\'a}zquez}, {Sengupta}, {Singh}, {Stanishev}, {Trager}, {Vazdekis}, {Viironen}, {Wild},
  {Zibetti}, \& {Ziegler}}]{califa1}
{S{\'a}nchez}, S.~F., {Kennicutt}, R.~C., {Gil de Paz}, A., {et~al.} 2012, \aap, 538, A8

\bibitem[{{Sato-Polito} {et~al.}(2019){Sato-Polito}, {Montero-Dorta}, {Abramo}, {Prada}, \& {Klypin}}]{SatoPolito2019}
{Sato-Polito}, G., {Montero-Dorta}, A.~D., {Abramo}, L.~R., {Prada}, F., \& {Klypin}, A. 2019, \mnras, 487, 1570

\bibitem[{{Sheth} {et~al.}(2001){Sheth}, {Mo}, \& {Tormen}}]{Sheth2001}
{Sheth}, R.~K., {Mo}, H.~J., \& {Tormen}, G. 2001, \mnras, 323, 1

\bibitem[{{Sheth} \& {Tormen}(1999)}]{ShethTormen1999}
{Sheth}, R.~K. \& {Tormen}, G. 1999, \mnras, 308, 119

\bibitem[{{Sheth} \& {Tormen}(2002)}]{ShethTormen2002}
{Sheth}, R.~K. \& {Tormen}, G. 2002, \mnras, 329, 61

\bibitem[{{Sheth} \& {Tormen}(2004)}]{Sheth2004}
{Sheth}, R.~K. \& {Tormen}, G. 2004, \mnras, 350, 1385

\bibitem[{{Sinha} \& {Garrison}(2020)}]{corrfunc}
{Sinha}, M. \& {Garrison}, L.~H. 2020, \mnras, 491, 3022

\bibitem[{{Springel}(2010)}]{Springel2010}
{Springel}, V. 2010, \mnras, 401, 791

\bibitem[{{Springel} {et~al.}(2018){Springel}, {Pakmor}, {Pillepich}, {Weinberger}, {Nelson}, {Hernquist}, {Vogelsberger}, {Genel}, {Torrey}, {Marinacci}, \& {Naiman}}]{Springel2018}
{Springel}, V., {Pakmor}, R., {Pillepich}, A., {et~al.} 2018, \mnras, 475, 676

\bibitem[{{Sunayama} {et~al.}(2022){Sunayama}, {More}, \& {Miyatake}}]{Sunayama2022}
{Sunayama}, T., {More}, S., \& {Miyatake}, H. 2022, arXiv e-prints, arXiv:2205.03277

\bibitem[{{Tempel} {et~al.}(2017){Tempel}, {Tuvikene}, {Kipper}, \& {Libeskind}}]{2017A&A...602A.100T}
{Tempel}, E., {Tuvikene}, T., {Kipper}, R., \& {Libeskind}, N.~I. 2017, \aap, 602, A100

\bibitem[{{Tucci} {et~al.}(2021){Tucci}, {Montero-Dorta}, {Abramo}, {Sato-Polito}, \& {Artale}}]{Tucci2021}
{Tucci}, B., {Montero-Dorta}, A.~D., {Abramo}, L.~R., {Sato-Polito}, G., \& {Artale}, M.~C. 2021, \mnras, 500, 2777

\bibitem[{{Vogelsberger} {et~al.}(2014{\natexlab{a}}){Vogelsberger}, {Genel}, {Springel}, {Torrey}, {Sijacki}, {Xu}, {Snyder}, {Bird}, {Nelson}, \& {Hernquist}}]{Vogelsberger2014b}
{Vogelsberger}, M., {Genel}, S., {Springel}, V., {et~al.} 2014{\natexlab{a}}, \nat, 509, 177

\bibitem[{{Vogelsberger} {et~al.}(2014{\natexlab{b}}){Vogelsberger}, {Genel}, {Springel}, {Torrey}, {Sijacki}, {Xu}, {Snyder}, {Nelson}, \& {Hernquist}}]{Vogelsberger2014a}
{Vogelsberger}, M., {Genel}, S., {Springel}, V., {et~al.} 2014{\natexlab{b}}, \mnras, 444, 1518

\bibitem[{{Walcher} {et~al.}(2014){Walcher}, {Wisotzki}, {Bekerait{\'e}}, {Husemann}, {Iglesias-P{\'a}ramo}, {Backsmann}, {Barrera Ballesteros}, {Catal{\'a}n-Torrecilla}, {Cortijo}, {del Olmo}, {Garcia Lorenzo}, {Falc{\'o}n-Barroso}, {Jilkova}, {Kalinova}, {Mast}, {Marino}, {M{\'e}ndez-Abreu}, {Pasquali}, {S{\'a}nchez}, {Trager}, {Zibetti}, {Aguerri}, {Alves}, {Bland-Hawthorn}, {Boselli}, {Castillo Morales}, {Cid Fernandes}, {Flores}, {Galbany}, {Gallazzi}, {Garc{\'\i}a-Benito}, {Gil de Paz}, {Gonz{\'a}lez-Delgado}, {Jahnke}, {Jungwiert}, {Kehrig}, {Lyubenova}, {M{\'a}rquez Perez}, {Masegosa}, {Monreal Ibero}, {P{\'e}rez}, {Quirrenbach}, {Rosales-Ortega}, {Roth}, {Sanchez-Blazquez}, {Spekkens}, {Tundo}, {van de Ven}, {Verheijen}, {Vilchez}, \& {Ziegler}}]{califa3}
{Walcher}, C.~J., {Wisotzki}, L., {Bekerait{\'e}}, S., {et~al.} 2014, \aap, 569, A1

\bibitem[{{Wang} {et~al.}(2022){Wang}, {Mao}, {Zentner}, {Guo}, {Lange}, {van den Bosch}, \& {Mezini}}]{Wang2022}
{Wang}, K., {Mao}, Y.-Y., {Zentner}, A.~R., {et~al.} 2022, \mnras, 516, 4003

\bibitem[{{Wechsler} {et~al.}(2006){Wechsler}, {Zentner}, {Bullock}, {Kravtsov}, \& {Allgood}}]{Wechsler2006}
{Wechsler}, R.~H., {Zentner}, A.~R., {Bullock}, J.~S., {Kravtsov}, A.~V., \& {Allgood}, B. 2006, \apj, 652, 71

\bibitem[{{White} \& {Frenk}(1991)}]{White1991}
{White}, S.~D.~M. \& {Frenk}, C.~S. 1991, \apj, 379, 52

\bibitem[{{White} \& {Rees}(1978)}]{White1978}
{White}, S.~D.~M. \& {Rees}, M.~J. 1978, \mnras, 183, 341

\bibitem[{{Yang} {et~al.}(2005){Yang}, {Mo}, {van den Bosch}, \& {Jing}}]{Yang2005}
{Yang}, X., {Mo}, H.~J., {van den Bosch}, F.~C., \& {Jing}, Y.~P. 2005, \mnras, 356, 1293

\end{thebibliography}

\end{document}